\documentclass[aps,                % American Physical Society (APS) styling
               prc,                % customize styling for APS journals
               showpacs,           % Control display of PACS
               notitlepage,
               onecolumn,          % One-column formatting
               10pt,               % 10 pt text size
               twoside,            % Two side paper
               floatfix,           % Avoid float problems when twocolumn is used
               superscriptaddress] % Associate authors with applications via superscript numbers
               {revtex4-1}

\pdfoutput=1
\usepackage{mathptmx}
\usepackage[T1]{fontenc}
\usepackage{geometry}         % To modify page geometry (margin, text width, ...)
\usepackage{fancyhdr}         % To modify headlines and footlines
\usepackage{graphicx}         % For pictures and cartoons
\usepackage[svgnames]{xcolor} % To color text
\usepackage{hyperref}         % To add URL and link
\usepackage{amsmath}          % Various features to facilitate writing math formula
\usepackage{amssymb}          % Various mathematical symbols
\usepackage{amsfonts}         % Various mathematical font (\math(bb)(bf)...)
\usepackage{amstext}          % To ute normal text in formula (\text)
\usepackage{textcomp}         % For Copyright
\usepackage{slashed}
\hypersetup{colorlinks=true,linkcolor=Blue,citecolor=Blue,urlcolor=Blue}

\DeclareMathAlphabet{\mathcal}{OMS}{cmsy}{m}{n}

\newcommand{\bs}[1]{p\!\!\!\!\! p \!\!\!\!\! p\!\!\!\!\! p}
\newcommand{\bsp}[1]{p\!\!\!\!\! p \!\!\!\!\! p\!\!\!\!\! p}
\newcommand{\bsq}[1]{q\!\!\!\!\! q}
\newcommand{\bsA}[1]{A\!\!\!\!\! A}
\newcommand{\bsP}[1]{P\!\!\!\!\! P}
\newcommand{\bsAA}[1]{A\!\!\!\!\!\!\! A}
\newcommand{\bsa}[1]{a\!\!\!\!\! a}
\newcommand{\bsu}[1]{u\!\!\!\!\! u}
\newcommand{\bsbt}[1]{\beta\!\!\!\!\!\! \beta \!\!\!\!\!\! \beta \!\!\!\!\!\! \beta}

\numberwithin{equation}{section}                                % number equations to the section

\fancypagestyle{firststyle}
{%
  \fancyhf{}
  \fancyhead[C]{PHYSICAL REVIEW C {\bf 87}, 034912 (2013)} % Left Odd
  \fancyfoot[L]{\small 0556-2813/2013/87(1)/034912(25)} % Left
  \fancyfoot[C]{\small 034912-\thepage} % Centered number of the page in footline
  \fancyfoot[R]{\small \textcopyright 2013 American Physical Society} % Right
}

\begin{document}

\title{\texorpdfstring{\begin{minipage}[c]{\textwidth}\centering Heavy quark scattering and quenching in a QCD medium at finite temperature and chemical potential \end{minipage}}{Heavy quark scattering and quenching in a QCD medium at finite temperature and chemical potential}}

\author{H.~Berrehrah}
\email{berrehrah@fias.uni-frankfurt.de}
\affiliation{\begin{minipage}[c]{\textwidth}Frankfurt Institute for Advanced Studies and Institute for Theoretical Physics, Johann Wolfgang Goethe Universit\"at, Ruth-Moufang-Strasse 1,\end{minipage}\\
60438 Frankfurt am Main, Germany \\ \vspace{0.5mm}}
%\affiliation{\begin{minipage}[c]{\textwidth}Institut for Theoretical Physics, Johann Wolfgang Goethe Universität, Max-von-Laue-Str. 1, 60438 Frankfurt am Main, Germany \end{minipage} \\ \vspace{0.5mm}}

\author{E.~Bratkovskaya}
\email{brat@th.physik.uni-frankfurt.de}
\affiliation{\begin{minipage}[c]{\textwidth}Frankfurt Institute for Advanced Studies and Institute for Theoretical Physics, Johann Wolfgang Goethe Universit\"at, Ruth-Moufang-Strasse 1,\end{minipage}\\
60438 Frankfurt am Main, Germany \\ \vspace{0.5mm}}
%\affiliation{\begin{minipage}[c]{\textwidth}Institut for Theoretical Physics, Johann Wolfgang Goethe Universität, Max-von-Laue-Str. 1, 60438 Frankfurt am Main, Germany \end{minipage} \\ \vspace{0.5mm}}

\author{W.~Cassing}
\email{wolfgang.cassing@theo.physik.uni-giessen.de}
\affiliation{\begin{minipage}[c]{\textwidth}Institut für Theoretische Physik, Universit\"at Giessen, 35392 Giessen, Germany\end{minipage} \\\vspace{0.5mm}}

\author{P.B.~Gossiaux}
\email{gossiaux@subatech.in2p3.fr}
\affiliation{\begin{minipage}[c]{0.98\textwidth}Subatech, UMR 6457, IN2P3/CNRS, Universit\'e de Nantes, \'Ecole des Mines de Nantes, 4 rue Alfred Kastler, 44307 Nantes cedex 3, France\end{minipage} \\ \vspace{1.5mm} %\normalfont{(Received 15 July 2013; revised 5 August 2013; published 28 August 2013)}
\vspace{2.5mm}}

\author{J.~Aichelin}
\email{aichelin@subatech.in2p3.fr}
\affiliation{\begin{minipage}[c]{0.98\textwidth}Subatech, UMR 6457, IN2P3/CNRS, Universit\'e de Nantes, \'Ecole des Mines de Nantes, 4 rue Alfred Kastler, 44307 Nantes cedex 3, France\end{minipage} \\ \vspace{1.5mm} %\normalfont{(Received 15 July 2013; revised 5 August 2013; published 28 August 2013)}
\vspace{2.5mm}}

% PACS, the Physics and Astronomy Classification Scheme
\pacs{24.10.Jv, 02.70.Ns, 12.38.Mh, 24.85.+p%
%\hfill DOI:\href{http://link.aps.org/doi/10.1103/PhysRevC.89.034912}{10.1103/PhysRevC.89.034912}
}

\begin{abstract}

  The heavy quark collisional scattering on partons of the quark gluon plasma (QGP) is studied in a QCD medium at finite temperature and chemical potential. We evaluate the effects of finite parton masses and widths, finite temperature $T$ and quark chemical potential $\mu_q$ on the different elastic cross sections for dynamical quasi-particles (on- and off-shell particles in the QGP medium as described by the dynamical quasi-particles model ``DQPM'') using the leading order Born diagrams. Our results show clearly the decrease of the $qQ$ and $gQ$ total elastic cross sections when the temperature and the quark chemical potential increase. These effects are amplified for finite $\mu_q$ at temperatures lower than the corresponding critical temperature $T_c (\mu_q)$. Using these cross sections we, furthermore, estimate the energy loss and longitudinal and transverse momentum transfers of a heavy quark propagating in a finite temperature and chemical potential medium. Accordingly, we have shown that the transport properties of heavy quarks are sensitive to the temperature and chemical potential variations. Our results provide some basic ingredients for the study of charm physics in heavy-ion collisions at Beam Energy Scan (BES) at RHIC and CBM experiment at FAIR.

\end{abstract}


\maketitle

  %---------------------------------------------------------------------------------------------------------------------------------------------
\section{Introduction}
\label{Introduction}
%---------------------------------------------------------------------------------------------------------------------------------------------
%\thispagestyle{firststyle}
\vskip -2mm

  The exploration of the phase diagram of strongly interacting matter is a major field of modern high-energy physics. The transition from hadronic to partonic degrees of freedom  at high temperatures or high baryon densities is one of the most interesting challenges of relativistic heavy-ion physics. The discovery of this phase transition is expected to elucidate some of the fundamental aspects of Quantum Chromo Dynamics (QCD), i.e.  confinement and chiral symmetry breaking. Moreover, the different phases of the QCD phase diagram play an important role in the evolution of the early universe and the structure of the core of neutron stars \cite{Weber:2000xd}.

  The transition from a hadronic to a partonic medium at small net-baryon densities is known to be a crossover. On the other hand, at high baryon densities one expects new phases of strongly interacting matter \cite{Wilczek:1999ym}. In particular, a first-order  deconfinement phase transition with a critical endpoint or a chiral symmetry restoration without deconfinement -- leading to a quarkyonic phase -- may occur at larger baryon densities.

  Recent  results from a statistical analysis of particle ratios measured in Pb+Pb and Au+Au collisions at SchwerIonen-Synchrotron (SIS), Alternating Gradient Synchrotron (AGS), Super Proton Synchrotron (SPS)  and Relativistic Heavy-Ion Collider (RHIC) energies point in this direction \cite{BraunMunzinger:2000mv,BraunMunzinger:2001ip,Becattini:2003wp,Averbeck:1998jb,Cleymans:1998fq}. The phase boundary between quark-gluon and hadronic  matter and the location of the (possible) critical endpoint is suggested by lattice QCD calculations \cite{Fodor:2004nz,Ejiri:2003dc} to occur for values of $\mu_B$ larger than about 400 MeV;  for $\mu_B \approx 3 \mu_q$ smaller than 400 MeV one expects a smooth cross over from the hadronic to the partonic phase. Besides this hypothetical first order transition a rich phase structure might occur at high baryon chemical potential $\mu_B$ or quark chemical potential $\mu_q$.
  
  Previous studies have shown that the high density regime of QCD is accessible with heavy-ion collisions at moderate collisions energies \cite{Randrup}. This is supported by the study of the freeze-out points from hadron-gas models as a function of the temperature and chemical potential \cite{Andronic:2008gu} or hadronic transport models like the hadron-string-dynamics (HSD) approach from \cite{Cassing:1999PR,Cassing:2000vx}, where    the highest baryon densities are predicted for moderate collisions energies.

  Recent progress in the exciting field of QCD at high baryon densities is driven by new experimental data from the Beam Energy Scan (BES) program at RHIC ($\sqrt{s_{NN}}$=7-200 GeV) and NA61 at the SPS ($\sqrt{s_{NN}}$=6.4-17.4 GeV). In future experiments of the  Compressed Baryonic Matter (CBM) collaboration at FAIR ($\sqrt{s_{NN}}$=2.7-8.3 GeV) and the Multi-Purpose Detector (MPD) at NICA ($\sqrt{s_{NN}}$=4-11 GeV) will provide additional information. The aim of all these experiments is to explore the QCD phase diagram at high net baryon densities and moderate temperatures in nucleus-nucleus collisions. This approach is complementary to the studies of matter at high temperatures and low net baryon densities performed at RHIC and the Large Hadron Collider (LHC) which are designed to study the properties of the deconfined medium at the highest available energy densities (and temperatures). Since lattice QCD (lQCD) does not provide robust results in this regime  one has to rely on effective QCD models that match lQCD for $\mu_q = 0$. Furthermore,
one has to look at the characteristics of the medium systematically in terms of collision energy and system size, by studying the strangeness, charm, collective flow and fluctuations possibly on an event-by-event basis.

  Charm physics is one of the promising signals that can be studied at upper FAIR energies.  Indeed, both hidden and open charm are expected to contribute to the total charm production also in this energy range. The total charm cross section experimentally is not well known close to threshold and the predictions for A+A rely on parametrisations of experimental data and imply large uncertainties towards threshold. Note that also perturbative QCD (pQCD) calculations show large uncertainties \cite{Frawley:2008kk}. Therefore, close to the kinematic production thresholds, an unknown territory has to be systematically explored \cite{Linnyk:2006ti,Linnyk:2007ti}. Furthermore, experimental data close to threshold for the elementary charm cross sections have to be taken before more robust conclusions can be expected for p+A and A+A collisions at FAIR/NICA energies. Note that there are presently no p+N data below 20 GeV and no A+A data below top SPS energies for charmonia.

  The anomalous suppression of charmonium due to screening effects in the Quark-Gluon Plasma (QGP) was predicted to be an experimental signal of the QGP formation by Matsui and Satz \cite{Matsui:1986dk} since particles containing heavy quarks like charm are produced predominantly in the early stage of the collision due to high kinematical thresholds. Heavy-flavour physics at FAIR thus aims to explore how charm is produced at beam energies close to the kinematical thresholds and how  charm propagates in hot and cold nuclear matter as well as in the partonic phase. Besides, the production mechanisms of $D$ and $J/\psi$ mesons will be sensitive to the conditions inside the early fireball.

  The scenarios for charm production in A+A collisions based on either hadronic \cite{Cassing:2000Exi,Cassing:2000vx} or partonic (statistical hadronization model, SHM) \cite{Andronic:2007zu} models have given different predictions. Especially the ratio of hidden to open charm ($J/\psi/D$) from the hadronic HSD model \cite{Linnyk:2006ti,Linnyk:2007ti} differs substantially from the one  in the partonic SHM model  \cite{Andronic:2007zu} since the $J/\psi/D$ ratio depends on the energy in the c.m.s. ($\sqrt{s}$) in the first model and is $\sqrt{s}$ independent in the second one. Besides, the $J/\psi/D$ ratio is about one order of magnitude higher in the hadronic than in partonic production scenarios which is due to the much lower threshold for $J/\psi$ production in N+N collisions than for $D + \bar{D}$ pairs. Therefore, the charm production is sensitive to the phases of matter and the ratio of hidden to open charm appears as a very promising probe for the production of charm and its propagation in the medium. In addition the collective flow of charm is expected to provide valuable information on the interaction strength of charm with its medium being of hadronic or partonic nature. Furthermore, it is generally questioned that charm degrees of freedom might achieve a chemical equilibrium in the fireball such that microscopic transport approaches are mandatory to shed some light on the nonlinear charm dynamics.

   With the future aim to implement the charm dynamics non-perturbatively into the Parton-Hadron-String Dynamics (PHSD) approach we have to specify the interaction cross sections of the charm degrees of freedom ($Q$) with the light partonic degrees of freedom incorporated in PHSD, i.e. dressed quarks, antiquarks ($q , {\bar q}$) and gluons ($g$). Accordingly, in this study we will compute the off-shell cross sections for the reactions $q Q \rightarrow q Q$ and $g Q \rightarrow g Q$ taking into account the quasi particle nature of the quarks and gluons at finite $T$ and $\mu_q$ in PHSD which are adopted from the dynamical quasi-particle model (DQPM) \cite{PhysRevD.70.034016,0954-3899-31-4-046,Wcassing2009EPJS}. These cross sections are then used to evaluate the energy and momentum losses of a heavy quark as a function of heavy-quark momentum, temperature and quark chemical potential. Our results finally will provide the basic ingredients for the microscopic study of charm physics for the  Beam Energy Scan (BES) program at RHIC and the future CBM experiment at FAIR. For a first step in this direction we refer the reader to Ref. \cite{Berrehrah:2014kba} where some of the quantities have already been evaluated at $\mu_q = 0$ at finite temperatures relevant at RHIC or LHC energies.

  The paper is organized as follows:  We first present in section \ref{DQP-finiteTmu} the basic ingredients needed for the heavy quark scattering in a finite temperature $T$ and quark chemical potential $\mu_q$ medium. Therefore, we fix the coupling constant, the parton masses and spectral functions and the gluon and fermion propagators as given by the DQPM at finite $T$ and $\mu_q$. The analysis of the on- and off-shell kinematics and the calculations of the on- and off-shell elastic cross sections of the scattering of a heavy quark ($Q$) and light quark ($q$)  and gluon ($g$) in a partonic medium at finite $T$ and $\mu_q$ are specified in Section \ref{qgQProcess}. In Section \ref{HQRates}, we calculate the interaction rate of the heavy quark in a such medium. Furthermore, we will perform  a quantitative analysis of the heavy-quark energy loss (sec.\ref{HQELoss}) and momentum loss (sec. \ref{HQPLoss}). Throughout Sections \ref{HQRates}-\ref{HQPLoss} we will point out the effects of finite masses and width, finite temperature and chemical potential on the heavy-quark transport properties. In Section \ref{summary} we summarize the main results and point out their future applications. 
  %---------------------------------------------------------------------------------------------------------------------------------------------
\section{Dynamical Quasi-Particles at finite temperature and chemical potential}
\label{DQP-finiteTmu}
%---------------------------------------------------------------------------------------------------------------------------------------------
\vskip -2mm

  The scattering of heavy quarks in vacuum and in a QGP medium in lowest-order QCD perturbation theory (pQCD) has extensively been studied in the literature \cite{PhysRevD.17.196,Combridge1979429}. Recent developments reconsidered the concept of perturbatively interacting massless quarks and gluons as constituents of the QGP, which scatter according to the leading (Born) diagrams. The treatment of non-perturbative effects in heavy quark scattering was first carried out by Braaten et al. \cite{PhysRevD.44.R2625,PhysRevD.44.1298,PhysRevD.44.2774} in thermal perturbation theory, denoted as hard thermal loop (HTL) approach and later by Peshier et al. \cite{Peshier:2006hi,Peshier:2008bg,Peigne:2008nd} and Gossiaux et al. \cite{Gossiaux:2004qw,PhysRevC.78.014904,Gossiaux2009203c,Gossiaux:2009hr,Gossiaux:2009mk,Gossiaux:2010yx,Gossiaux:2006yu}. In Ref.\cite{Berrehrah:2013mua} we considered all the effects of the non-perturbative nature of the strongly interaction quark-gluon plasma (sQGP) constituents, i.e. the large coupling, the multiple scattering etc., where we refrain from a fixed-order thermal loop calculation relying on perturbative self-energies (calculated in the limit of infinite temperature) to fix the in-medium masses of the quarks and gluons and pursue instead a more phenomenological approach. The multiple strong interactions of quarks and gluons in the sQGP are encoded in their effective propagators with broad spectral functions. The effective propagators, which can be interpreted as resummed propagators in a hot and dense QCD environment, have been extracted from lattice data in the scope of the DQPM \cite{Wcassing2009EPJS,Cassing:2009vt,Cassing2007365}.

  In this work we extend our study of the scattering processes $q Q \rightarrow q Q$ and $g Q \rightarrow g Q$ in Refs. \cite{Berrehrah:2013mua,Berrehrah:2014kba} for a partonic medium at finite temperature $T$ and in particular at finite chemical potential $\mu_q$. The gluons ($g$) and light ($q$) and heavy ($Q$) quark masses as well as the fundamental ingredients (infrared regulator 'IR' and running coupling '$\alpha_s$') involved in the scattering amplitudes are determined in the framework of the DQPM. The dependencies of these quantities on $T$ and $\mu_q$  will be discussed below.

%------------------------------------------------------------------------------------------------------------------------------------------------------------------------------
\subsection[Dynamical Quasi-Particles at finite $T$ and quark chemical potential $\mu_q$]{Dynamical Quasi-Particles at finite temperature and quark chemical potential $\mu_q$} %------------------------------------------------------------------------------------------------------------------------------------------------------------------------------

  The DQPM describes QCD properties in terms of  single-particle Green's functions (in the sense of a two-particle irreducible (2PI) approach) and leads to the notion of the constituents of the sQGP being strongly interacting massive effective quasi-particles with broad spectral functions (due to the high interaction rates). The strategy for the determination of parton masses and widths within the DQPM approach is to fit the analytical expression of the dynamical quasi-particle entropy density $s^{DQP}$ to the lQCD entropy density \emph{``$s^{lQCD}$''} which allows to fix the few parameters present in $s^{DQP}$ by lattice data in equilibrium \cite{PhysRevD.70.034016,0954-3899-31-4-046,Cassing2007365,Wcassing2009EPJS}.

  The variation of parton masses as a function of the medium properties is described by the spectral functions which are (except of a factor) identical to the imaginary part of the retarded propagator. These are no longer $\delta-$ functions in the invariant mass squared (as in the case for bare masses) \cite{Cassing:2009vt}. For the current analysis,  we use the approximation of momentum-independent real and imaginary parts of the retarded self-energy, which are - for a given temperature $T$ - proportional to the parton mass and width, respectively \cite{Wcassing2009EPJS}. A partonic propagator $\Delta$ is expressed in the Lehmann representation in terms of the spectral function $\Delta (p) = \int \frac{d \omega}{2 \pi} \frac{A (\omega, \bs{p})}{p_0 -\omega}$.  An often used ansatz to model a non-zero width is obtained by replacing the free spectral function $A_0 (p) = 2 \pi [\delta (\omega - p^2)^2 - \delta (\omega + p^2)^2]$ by a Lorentzian form \cite{Wcassing2009EPJS,Berrehrah:2013mua} which also emerges from Kadanoff-Baym theory in first order gradient expansion.

  The extension of the DQPM to finite quark chemical potential $\mu_q$ is more delicate since a guidance by lQCD is presently very limited. In the simple quasiparticle model one may use the stationarity of the thermodynamic potential with respect to self-energies and (by employing Maxwell relations) derive a partial differential equation for the coupling $g^2(T,\mu_q)$ which may be solved with a suitable boundary condition for $g^2(T,\mu_q=0)$ \cite{Peshier:1995ty,Levai:1997yx,Peshier:1999ww,Peshier:2002ww}. Once $g^2(T,\mu_q)$ is known one can evaluate the changes in the quasiparticle masses with respect to $T$ and $\mu_q$, i.e. $\partial M_x^2/\partial \mu_q$ and $\partial M_x^2/\partial T$ (for $x=g,q,\bar{q}$) and calculate the change in the 'bag pressure' (cf. Refs. \cite{Peshier:1995ty,Levai:1997yx,Peshier:1999ww,Peshier:2002ww,Rebhan:2003wn} for details). However, such a strategy cannot be taken over directly in the DQPM since additionally the quasiparticle widths $\gamma_x(T,\mu_q)$ have to be known in the $(T,\mu_q)$ plane in this case.

 In hard-thermal-loop (HTL) approaches \cite{Braaten:1991gm,Blaizot:1993be} the damping of a light quark (or gluon) depends weakly on the quark chemical potential explicitly \cite{Vija:1994is}. This, however, has to be considered with care since HTL approaches primarily address Landau damping and assume small couplings $g^2$. Accordingly, these concepts should be applied at sufficiently high temperature, only. Present lQCD calculations suggest that the ratio of pressure to energy density, $P/\epsilon$, is approximately independent of $\mu_q$ as a function of the energy density $\epsilon$ \cite{Fodor:2002km}. The functional dependence of the quasiparticle width $\gamma$ on $T$ and $\mu_q$ thus has to be modeled in line with 'lattice phenomenology'. We proceed in this paper to some scaling hypothesis in order to extend the definition of DQPM masses and widths to a finite chemical potential \cite{Wcassing2009EPJS}. Assuming three light flavors ($q= u,d,s)$ and all chemical potentials to be equal ($\mu_u = \mu_d = \mu_s = \mu_q$) the gluon and light quark masses are taken for $T$ and $\mu_q$ as

{\setlength\arraycolsep{-6pt}
\begin{eqnarray}
\label{equ:Sec2.6}
& & M_g^2 (T, \mu_q) = \frac{g^2 (T/T_c)}{6} \Biggl((N_c + \frac{1}{2} N_f) T^2 + \frac{3}{2} \sum_q \frac{\mu_q^2}{\pi^2} \Biggr),
\nonumber\\
& & {} M_q^2 (T, \mu_q) = \frac{N_c^2 - 1}{8 N_c} g^2 (T/T_c) \Biggl( T^2 + \frac{\mu_q^2}{\pi^2} \Biggr).
\end{eqnarray}}

  This functional form is inspired by HTL masses but slightly differs from those in order to incorporate an explicit scaling with the effective temperature for $N_f = N_c = 3$

\begin{equation}
\label{equ:Sec2.7} T^{*2} = T^2+\frac{\mu_q^2}{\pi^2} ,
\end{equation}

which implies no additional parameter in the DQPM. Thus $M_g^2 (T, \mu_q)$ and $M_q^2 (T, \mu_q)$ are given by

{\setlength\arraycolsep{-6pt}
\begin{eqnarray}
\label{equ:Sec2.7bis}
& & M_g^2 (T^{\star}, \mu_q) = \frac{g^2 (T^{\star}/T_c(\mu_q))}{6} (N_c + \frac{1}{2} N_f) \ {T^{\star}}^2,
\nonumber\\
& & {} M_q^2 (T^{\star}, \mu_q) = \frac{N_c^2 - 1}{8 N_c} g^2 (T^{\star}/T_c(\mu_q)) \ {T^{\star}}^2,
%\\
%& & {}  \gamma_g (T) = \frac{1}{3} N_c \frac{g^2 (T/T_c) T}{8 \pi} \ln \left(\frac{2 c}{g^2 (T/T_c)} + 1 \right),
%\nonumber\\
%& & {} \gamma_q (T) = \frac{1}{3} \frac{N_c^2 - 1}{2 N_c} \frac{g^2 (T/T_c) T}{8 \pi} \ln \left(\frac{2 c}{g^2 (T/T_c)} + 1 \right).
%\nonumber
\end{eqnarray}}

  with the coupling constant $g^2 (T^{\star}/T_c (\mu_q))$ in (\ref{equ:Sec2.7bis}) is considered here as depending on the medium temperature and for $T^{\star} > T_s$ is given by:
{\setlength\arraycolsep{0pt}
\begin{eqnarray}
\label{equ:Sec2.8bis}
& & \hspace{-0.1cm} \displaystyle{g^2 (T^{\star}, \mu_q) = \frac{48 \pi^2}{(11 N_c - 2 N_f) \ln \left(\lambda^2 (\frac{T^{\star}}{T_c(\mu_q)} - \frac{T_s}{T_c(\mu_q)})^2 \right)}}. %\hspace{0.3cm}  T > T^{\star} = 1.19 \ T_c,
%%\nonumber\\
%% & {} \hspace{-0.1cm} \displaystyle{g^2 (T/T_c) \rightarrow g^2 (T^{\star}/T_c)
%%left(\frac{T^{\star}}{T} \right)^{3.1}  \hspace{0.7cm}  T < T^{\star} = 1.19 \ T_c.}
\end{eqnarray}}
  Since the coupling (squared) in the DQPM is a function of $T/T_c$ a straight forward extension of the DQPM to finite $\mu_q$ is to consider the coupling as a function of $T^*/T_c(\mu_q)$ with a $\mu_q$-dependent critical temperature,

\begin{equation}
\label{equ:Sec2.8} \frac{T_c(\mu_q)}{T_c(\mu_q=0)} =  \sqrt{1 -  \alpha \ \mu_q^2} \approx 1 - \alpha/2 \ \mu_q^2 + \cdots
\end{equation}

with $\alpha \approx$ 8.79 GeV$^{-2}$. The expression of $T_c(\mu_q)$ in (\ref{equ:Sec2.8}) is obtained by requiring a constant energy density $\epsilon$ for the system at $T=T_c(\mu_q)$ where $\epsilon$ at $T_c(\mu_q = 0) \approx 0.158$ GeV is fixed by lattice QCD calculation at $\mu_q = 0$. The coefficient in front of the $\mu_q^2$-dependent part at first sight appears arbitrary but can be compared to recent lQCD calculations (for imaginary quark chemical potentials) at finite (but small) $\mu_B$ which gives \cite{Delia}

\begin{equation} \label{eli}
\frac{T_c(\mu_B)}{T_c} = 1 - \kappa \left( \frac{\mu_B}{T_c} \right) ^2 + \cdots
\end{equation} with $\kappa$ = 0.013(2). Rewriting (\ref{equ:Sec2.8}) in the form (\ref{eli}) and using $\mu_B \approx 3 \mu_q$ we get $\kappa_{DQPM} \approx 0.0122$ which compares very well with the lQCD result. Consequently one has to expect an approximate scaling of the DQPM results if the partonic width is assumed to have the form,

{\setlength\arraycolsep{-6pt}
\begin{eqnarray}
\label{equ:Sec2.9}
& & {}  \gamma_g (T,\mu_q) = \frac{1}{3} N_c \frac{g^2 (T^*/T_c(\mu_q))}{8 \pi} \, T \ \ln \left(\frac{2 c}{g^2 (T^*/T_c(\mu_q))} + 1 \right),
\nonumber\\
& & {} \gamma_q (T,\mu_q) = \frac{1}{3} \frac{N_c^2 - 1}{2 N_c} \frac{g^2 (T^*/T_c(\mu_q))}{8 \pi} \, T \ \ln \left(\frac{2 c}{g^2 (T^*/T_c(\mu_q))} + 1 \right).
\end{eqnarray}}

  This choice leads to an approximate independence of the potential energies per degree of freedom as a function of $\mu_q$.  Nevertheless, the conjecture (\ref{equ:Sec2.9}) should be explicitly controlled by future lQCD studies for $N_f$=3 at finite quark chemical potential. Unfortunately, this task is presently out of reach and one has to live with the uncertainty in (\ref{equ:Sec2.9}) which is assumed in the following investigations.

  We recall that within the scaling hypothesis (\ref{equ:Sec2.7})-(\ref{equ:Sec2.9}) the results for the masses and widths at finite $T$ stay about the same as a function of $T^*/T_c(\mu_q)$ when dividing by the temperature $T$ \cite{Wcassing2009EPJS}.

  For a finite quark chemical potential $\mu_q$ the energy density $\epsilon$ in the DQPM is seen to scale well with $(T/T_{c}(\mu_q=0))^4$ as a function of temperature for $T^*/T_c(\mu_q)> 3$, however, increases slightly with $\mu_q$ close to the phase boundary where the scaling is violated on the level of 20\%. This violation in the scaling is essentially due to an increase of the pressure $P$ \cite{Wcassing2009EPJS}. Note that a quark chemical potential of 0.21 GeV corresponds to a baryon chemical potential of $\mu_B \approx 3 \mu_q = $0.63 GeV which is already substantial and the validity of (\ref{equ:Sec2.8}) becomes questionable. Since the pressure $P$ is obtained from an integration of the entropy density $s$ over temperature $T$ the increase in $P$ with $\mu_q$ can directly be traced back to a corresponding increase in entropy density.

\begin{figure}[h!] %tbh!
%\begin{minipage}{0.5\textwidth}
\begin{center}
     \includegraphics[height=8.7cm, width=8.5cm]{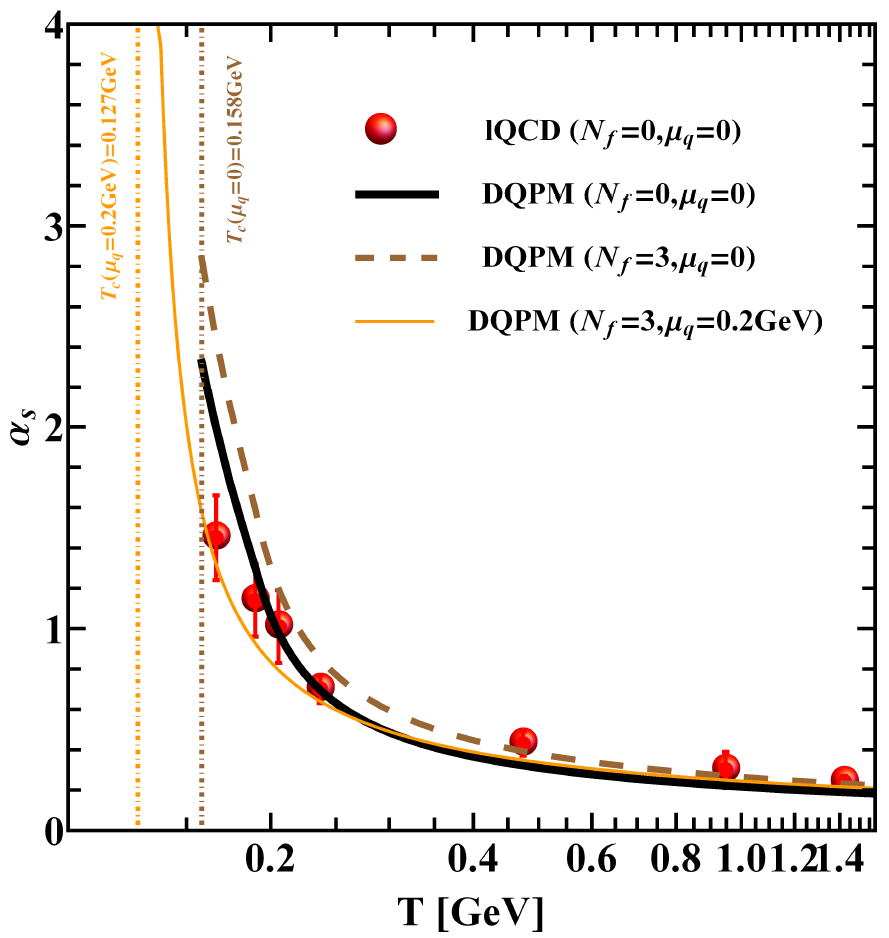} \hspace{5mm}
     \includegraphics[height=7.9cm, width=8.5cm]{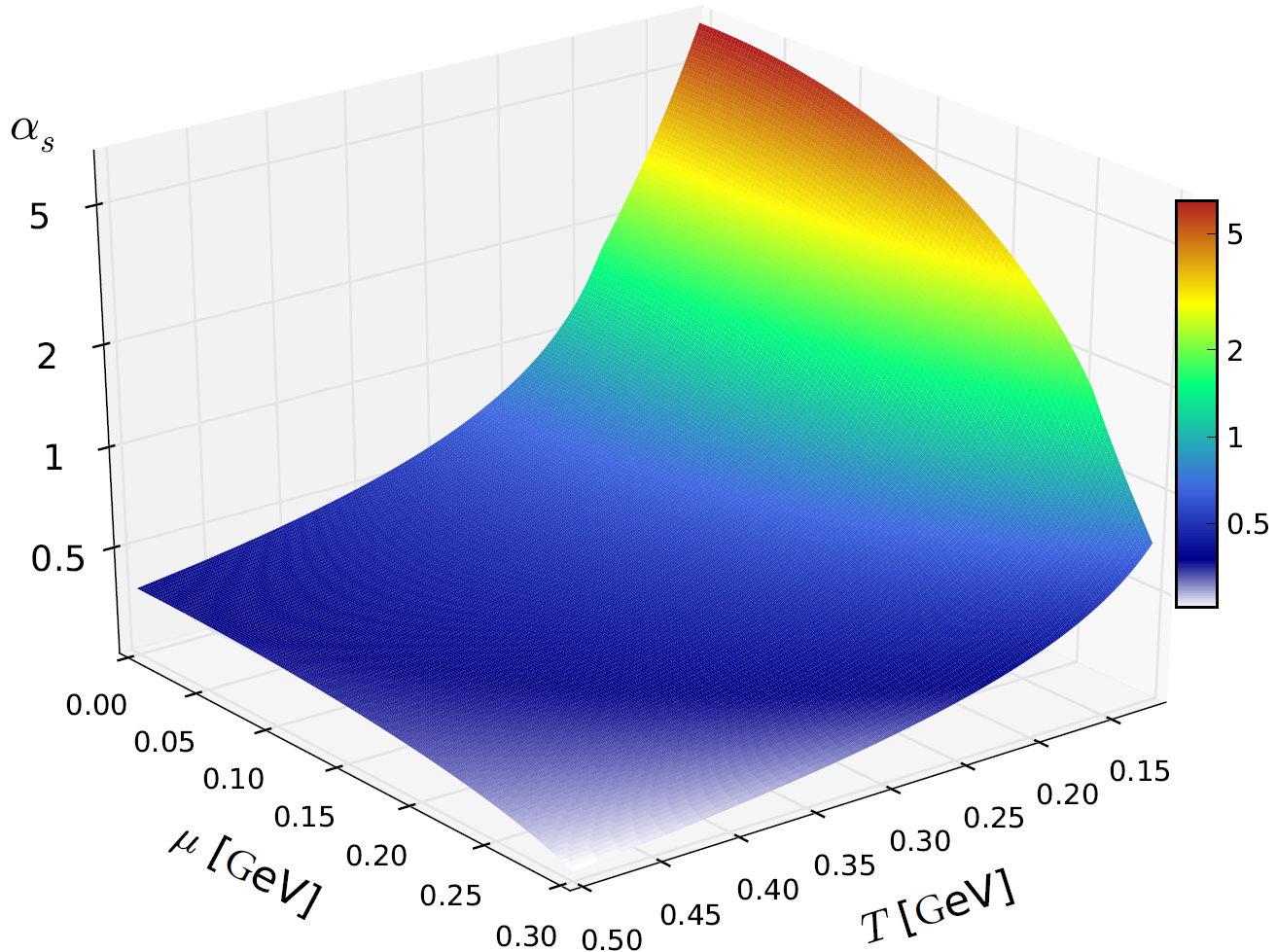} \hspace{5mm}
\end{center}
%\end{minipage}
%\begin{minipage}{0.45\textwidth}
\vspace{-4mm}
\caption{\emph{(Color online) (left) The DQPM running coupling  $\alpha_s (T,\mu_q) = g^2(T^{\star}/T_c(\mu_q))/(4 \pi)$ as a function of $T$ in the lQCD for $N_f = 0$ (red spheres) \cite{Kaczmarek:2004gv} and in the DQPM at zero chemical potential for $N_f = 0$ (black line) and $N_f = 3$ (dashed brown line). The result for the DQPM at finite quark chemical potential ($\mu_q = 0.2$ GeV) for $N_f = 3$ is given by the orange thin line, (right) 3D plot of $\alpha_s (T,\mu_q)$.}}
\label{fig:DQPM-alpha}
%\end{minipage}
\end{figure}

  Next we discuss the influence of a finite chemical potential $\mu_q$ on the running coupling $\alpha_s$, the DQPM masses, widths and parton spectral functions. The running coupling $\alpha_s = g^2/(4 \pi)$ is presented in Fig. \ref{fig:DQPM-alpha}-(a) as a function of the temperature $T$ for $\mu_q=0$ and in Fig. \ref{fig:DQPM-alpha}-(b) for finite $T$ and $\mu_q$. One sees that $\alpha_s$ is larger than 1 near $T_c (\mu_q)$ and non-perturbative effects are most pronounced at these temperatures, with $T_c (\mu_q=0)=0.158$ GeV and $T_c (\mu_q=0.2$GeV) $\approx 0.127$ GeV. Note that close to $T_c(\mu_q=0)$ the full coupling calculated on the lattice increases with decreasing temperature much faster than in the pQCD regime (at high $T$). A finite $\mu_q$ leads to a smaller value of the coupling constant as compared to the $\mu_q = 0$, except close to the corresponding critical temperature $T_c(\mu_q)$. Indeed, the value of $\alpha_s (T_c(\mu_q=0)) = 2.84$ while $\alpha_s (T_c(\mu_q=0.2$ GeV$))=4.7$. For temperatures larger than $T_c(\mu_q=0)$ one finds systematically $\alpha_s(T,\mu_q=0) > \alpha_s(T,\mu_q)$, as shown in Fig. \ref{fig:DQPM-alpha}-(b). 

\begin{figure}[h!] %tbh!
\begin{center}
  \includegraphics[height=8.3cm, width=8.7cm]{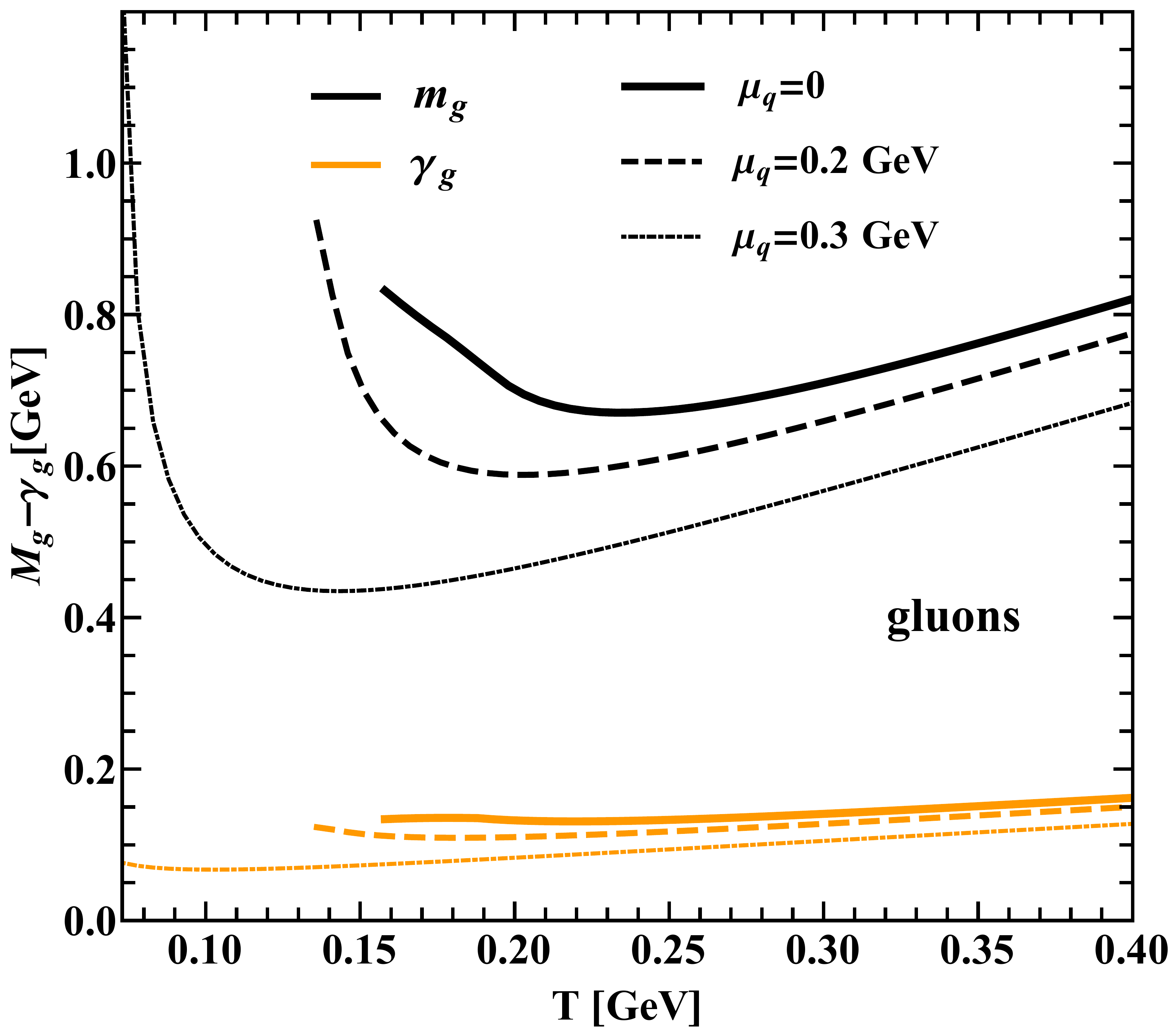}
  \includegraphics[height=8.3cm, width=8.7cm]{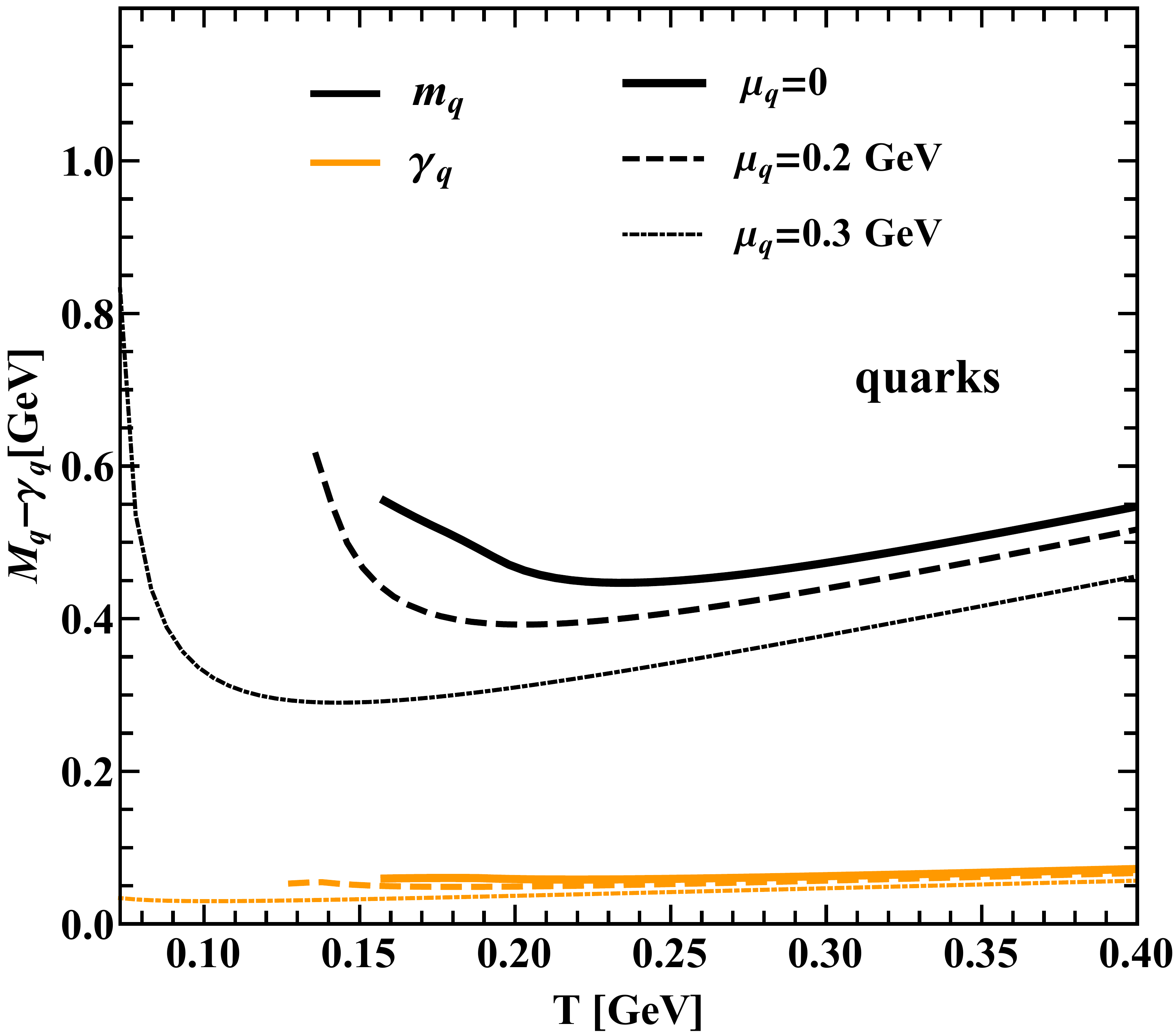}
\end{center}
\vspace{-4mm}
\caption{\emph{(Color online) The DQPM pole masses and widths for the gluons ($M_g$, $\gamma_g$) (left) and light quarks ($M_q$, $\gamma_q$) (right) given by (\ref{equ:Sec2.6}) and (\ref{equ:Sec2.9}) as a function of temperature for different values of the quark chemical potential $\mu_q$.}}
\label{fig:DQPM-MGamma}
\end{figure}

   The DQPM masses and widths for the gluon ($g$) and light quarks ($q$), given by (\ref{equ:Sec2.6}) and (\ref{equ:Sec2.9}), are presented in Fig. \ref{fig:DQPM-MGamma}-(a) and (b), respectively. Since the width is found to be much smaller as the pole mass, the excitations can well be considered as dynamical (off-shell) quasiparticles. For larger $T$, after a shallow minimum at $T \approx 1.2 T_c(\mu_q)$, the width $\gamma$ increases slowly with $T$ and even for  large $T$ is to a good accuracy proportional to the temperature (and also to the mass).

  In Figs. \ref{fig:DQPM-MGamma}-(a) and (b), the masses and widths are plotted as a function of the temperature. It is seen that the gluon and light quark masses and widths decrease at fixed $T$ with increasing quark chemical potential $\mu_q$. The finite $\mu_q$ has a larger effect on the masses as compared to the widths. For completeness we note: $T_c (\mu_q=0)=0.158$ GeV, $T_c (\mu_q=0.2$GeV$)=0.127$ GeV and $T_c (\mu_q=0.3$GeV$)=0.072$ GeV.

  Using the pole masses and widths (\ref{equ:Sec2.6}) and (\ref{equ:Sec2.9}) and the DQPM running coupling (cf. Fig.\ref{fig:DQPM-alpha}), the Breit-Wigner spectral functions for the different QGP species are completely determined, i.e. the imaginary parts of the retarded propagators. Accordingly, also the full retarded propagators for the effective partonic degrees of freedom are known in first order gradient expansion and we can proceed with the calculation of some amplitudes keeping in mind the uncertainties at high $\mu_q$.
  %---------------------------------------------------------------------------------------------------------------------------------------------
\section{$q Q$ and $g Q$ \ elastic scattering at finite $T$ and $\mu_q$}
\label{qgQProcess}
%---------------------------------------------------------------------------------------------------------------------------------------------
\vskip -2mm

   The matrix elements for the $q Q \rightarrow q Q$ and $g Q \rightarrow g Q$ channels have been calculated for the case of massless partons in the vacuum in Refs. \cite{PhysRevD.17.196,Combridge1979429}. In this section we study the $qQ$ and $gQ$ elastic scattering in the QGP medium at finite temperature $T$ and chemical potential $\mu_q$  considering the case of on- and off-shell gluons and light and heavy quarks. The partons are dressed by effective masses in the on-shell case and are dressed by the DQPM spectral functions with a finite width in the off-shell case. We refer to the on-shell study as DpQCD approach (Dressed pQCD) and to the off-shell case by IEHTL approach (Infrared Enhanced Hard Thermal Loop). For the DpQCD approach the $q Q$ and $g Q$ elastic cross section at finite $T$ and $\mu_q$ is determined by using (i)   the running coupling $\alpha_s (T, \mu_q)$ (Fig. \ref{fig:DQPM-alpha}), and (ii) the DQPM pole masses for the incoming and outgoing quarks and gluons. The DQPM gluon pole mass serves also as an infrared regulator in the gluon propagator.

   Considering in- and out- dynamical quasi-particles (DQP), the corresponding quasi-elastic IEHTL cross section $\sigma^{\textrm{IEHTL}}$ for the process $(1)^{m^{(1)}} + (2)^{m^{(2)}} \rightarrow (3)^{m^{(3)}} + (4)^{m^{(4)}}$ is deduced by the convolution of the modified pQCD cross section $\sigma$, where complex propagators are considered in the transition matrix elements, with the spectral functions, i.e.

\begin{widetext}
{\setlength\arraycolsep{0pt}
\begin{eqnarray}
%& & i) \ \sigma^{IEHTL} (s, m^{(1)}, m^{(2)}) = \int d m^{(3)} \ d m^{(4)} \ \sigma (s, m^{(1)}, m^{(2)}, m^{(3)}, m^{(4)}) \ A_{(3)} (m^{(3)}) \ A_{(4)} (m^{(4)})
%\\
%& & {} ii) \ \sigma^{IEHTL} (s, m^{(3)}, m^{(4)}) = \int d m^{(1)} \ d m^{(2)} \ \sigma (s, m^{(1)}, m^{(2)}, m^{(3)}, m^{(4)}) \ A_{(1)} (m^{(1)}) \ A_{(2)} (m^{(2)})
%\nonumber\\
& & {} \sigma^{\textrm{IEHTL}} (s) = \int d m^{(1)} \ d m^{(2)} \ d m^{(3)} \ d m^{(4)} \ \sigma (s, m^{(1)}, m^{(2)}, m^{(3)}, m^{(4)}) \rho_{(1)}^{\textrm{BW}} (m^{(1)}) \ \rho_{(2)}^{\textrm{BW}} (m^{(2)}) \ \rho_{(3)}^{\textrm{BW}} (m^{(3)}) \ \rho_{(4)}^{\textrm{BW}} (m^{(4)}),
%\nonumber
\label{equ:Sec3.1}
\end{eqnarray}}
\end{widetext}

 where $\rho_{(i)}^{\textrm{BW}} (m^{(i)})$ is the Breit-Wigner spectral function of the particle $i$, normalized as $\displaystyle \int_0^{\infty} d m^{(i)} \rho_{(i)}^{\textrm{BW}} (m^{(i)}) = 1$. The on- and off-shell cross sections for $qQ$, $gQ$ scattering in the sQGP at finite $T$ and $\mu_q$ are obtained using the DQPM parametrizations for the quark (gluon) self-energies, spectral functions and interaction strength at finite $T$ and $\mu_q$. Here, in principle,  a two-particle correlator should appear, but since we work in a 2PI motivated scheme the partons in the sQGP can be characterized by (dressed) single-particle propagators (cf. Ref. \cite{Berrehrah:2014kba}).

\begin{widetext}
\begin{figure}[h!]
\begin{center}
      \includegraphics[height=3.4cm, width=4cm]{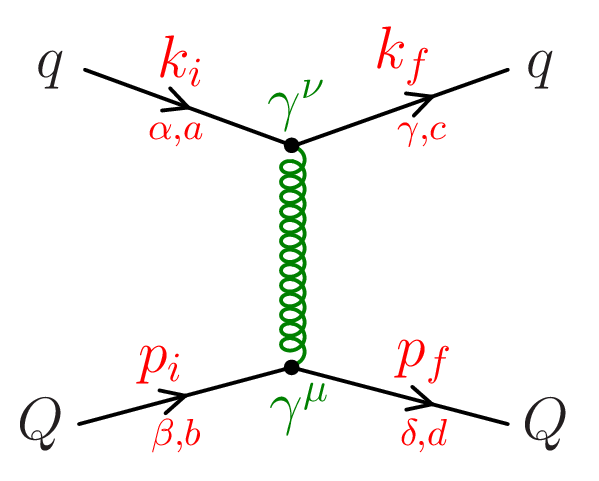}
      \hspace{0.8cm}
      \includegraphics[height=3.3cm, width=12.5cm]{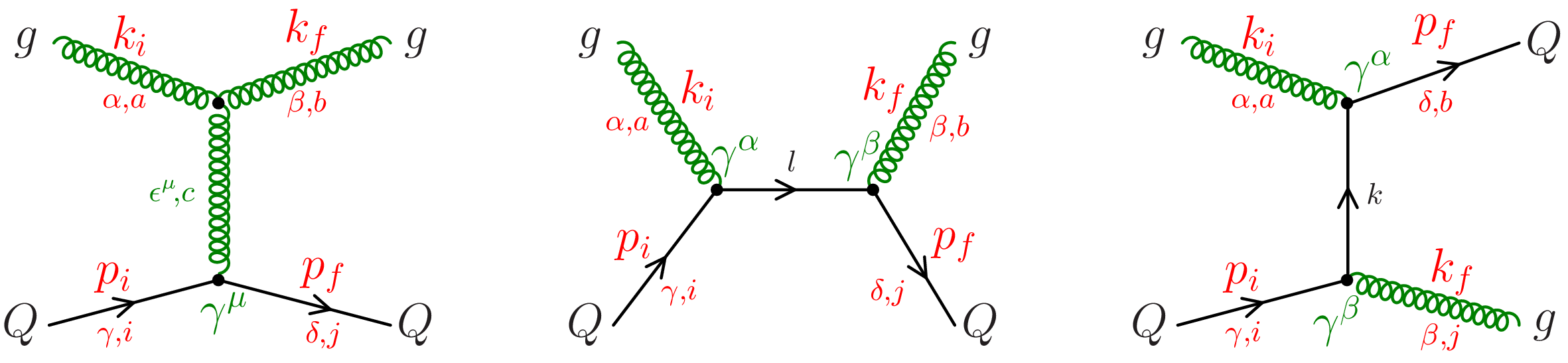}
\end{center}
\caption{\emph{(Color online) Feynman diagrams for the $q Q \rightarrow q Q$ and $g Q \rightarrow g Q$ scattering process. Latin (Greek) subscripts denote colour (spin) indices. $k_i$, resp $p_i$ ($k_f$, resp $p_f$) denote the initial (final) 4-momentum of the light quark or the gluon, resp the heavy quark. The invariant energy squared is given by $s=(p_i + k_i)^2$, $t=(p_i - p_f)^2$, $u=(p_i - k_f)^2$.} \label{fig:qgQFeynmanDiagram}}
\end{figure}
\end{widetext}

  In the context of the hot and dense QGP, the elementary Feynman diagrams for the $qQ$ and $gQ$ elastic scattering at order $O(\alpha_s)$ are illustrated in Fig.  \ref{fig:qgQFeynmanDiagram}.

%------------------------------------------------------------------------------------------------------------------------------------------------------------------------------
\subsection[On- and off-shell $qQ$ elastic scattering]{On- and off-shell $qQ$ elastic scattering} %------------------------------------------------------------------------------------------------------------------------------------------------------------------------------

  The process $q Q \rightarrow q Q$ is calculated here to lowest order in the perturbation expansion using the extended Feynman rules for massless quarks in Politzer's review \cite{DavidPolitzer1974129} for the case of finite masses and widths. The color sums are evaluated using the techniques discussed in Ref. \cite{DavidPolitzer1974129}; the spin sums will be discussed below. Contrary to the case of massless gluons where the \emph{``Transverse gauge''} is used, the \emph{``Lorentz covariance''} %\textcolor{red}{%please explain Lorentz covariance in this context a bit more}
  is used for the case of massive gluons here since a finite mass in the gluon propagator allows to fix the 0'th components of the gluon fields $A^0_a$ ($a=1, \cdots, 8$) by the spatial degrees of freedom  $A^k_a (k=1,2,3)$. Furthermore, the divergence encountered in the $t$-channel (Ref. \cite{PhysRevD.17.196,Combridge1979429,Combridge1977234}) -- when calculating the total cross sections $\sigma^{qQ}$ and $\sigma^{gQ}$ -- is cured self-consistently in the DpQCD and IEHTL models since the infrared regulator is given by the finite DQPM gluon mass (and width) in the DpQCD (IEHTL) model. For on-shell $qQ$ elastic scattering, the $t$-channel invariant squared amplitude - averaged over the initial spin and color degrees of freedom and summed over the final state spin and color - $\mathcal{M}_t$ is given by

\begin{widetext}
{\setlength\arraycolsep{0pt}
\begin{eqnarray}
\label{equ:Sec3.2}
\sum |\mathcal{M}_t|^2 & & \ = \frac{4 g^4}{9 (t - m_g^2)^2} \biggl[ (s-M_Q^2-m_q^2)^2 + (u - M_Q^2 - m_q^2)^2 + 2 (M_Q^2 + m_q^2)t \biggr],
\end{eqnarray}}
\end{widetext}

  where $m_q$ ($M_Q$) is the light quark (heavy quark) mass and $m_g$ is the DQPM exchanged gluon mass.

  In the off-shell picture we take into account not only the finite masses of the partons, but also their spectral functions, i.e. their finite widths. Since the light quark and heavy quark masses change before and after the scattering ('quasi-elastic' process) we introduce the mass $m_q^i$ for the initial $q$ and $m_q^f$ for the final $q$, and allow for different masses of the heavy quark, $M_Q^i$ for the initial $Q$ and $M_Q^f$ for the final $Q$. The squared amplitude -- averaged over the initial spin and color degrees of freedom and summed over the final state spin and color -- gives:

\begin{widetext}
{\setlength\arraycolsep{0pt}
\begin{eqnarray}
\label{equ:Sec3.3}
\sum |\mathcal{M}|^2 & & \ = \frac{2 g^4}{9 \bigl[(t-m_g^2)^2+4\gamma_g^2 q_0^2\bigr]} \biggl[ 4 \left(p_f^{\mu} p_i^{\nu} + p_i^{\mu} p_f^{\nu} + g^{\mu \nu} \frac{t}{2} \right) \biggr] \biggl[ 4 \left(k_{f,\mu} k_{i,\nu} + k_{i,\mu} k_{f,\nu} + g_{\mu \nu} \frac{t}{2} \right) \biggr],
\end{eqnarray}}
\end{widetext}

 where we have incorporated the DQPM propagators (i.e. $t_{\pm}^{*} = t - m_g^2 \pm 2 i \gamma_g q_0$, with $m_g$, $\gamma_g$ is, respectively, the effective gluon mass and total width at temperature $T$ and quark chemical potential $\mu_q$ and $q^0 = p_f^0 - p_i^0 = k_f^0 - k_i^0$ is the gluon energy in the $t$-channel). Thus the divergence in the gluon propagator in the $t$ channel is regularized.

  The relative contribution of the off-shell partons to the pQCD cross section is expected to change due to different kinematical thresholds and to the changes in the matrix element- corresponding to the diagram in Fig. \ref{fig:qgQFeynmanDiagram}. The off-shell kinematical limits for the momentum transfer squared $t$ and the expressions of the Mandelstam variables in the case of off-shell heavy quark scattering are given in Ref. \citep{Berrehrah:2013mua}.

%------------------------------------------------------------------------------------------------------------------------------------------------------------------------------
\subsection[On- and off-shell $gQ$ elastic scattering]{On- and off-shell $gQ$ elastic scattering} %------------------------------------------------------------------------------------------------------------------------------------------------------------------------------

  The invariant amplitudes for the three graphs (shown in Fig.\ref{fig:qgQFeynmanDiagram}) for the case of massive heavy quarks and massless gluons is given according to Combridge \cite{Combridge1979429} and revisited in \cite{Berrehrah:2013mua}. In Ref. \cite{Berrehrah:2013mua} we have already studied the scattering of massive on- and off- shell heavy quarks on massive on- and off-shell gluon in a finite temperature medium at $\mu_q=0$. In this paper we extend this study to a medium at finite temperature and chemical potential. Therefore, the lowest-order amplitude for the process $g Q \rightarrow g Q$, obtained from the Feynman rules of the gauge theory by the sum of the amplitudes of the three graphs (cf. appendix A.4 of \cite{Berrehrah:2013mua}), where in order to obtain the correct result for the squared matrix element

{\setlength\arraycolsep{0pt}
\begin{eqnarray}
<|\mathcal{M}|^2> = \frac{1}{4} \sum_{spins} T_{\alpha \beta} T_{\alpha' \beta'}^{\star} \epsilon_i^{\alpha} \epsilon_i^{\star \alpha'} \epsilon_f^{\beta} \epsilon_f^{\star \beta'},
%\nonumber\\
%& & {}
\label{equ:Sec3.4}
\end{eqnarray}}

  we have to use appropriate projection operators for the transverse polarisation states

{\setlength\arraycolsep{-5pt}
\begin{eqnarray}
\label{equ:Sec3.5}
& & \sum_{pol, i} \epsilon_{i,\alpha} \epsilon_{i, \alpha' } =  g_{\alpha \alpha'} - \frac{k_{i, \alpha} \ k_{i, \alpha'}}{(m_g^i)^2},
\nonumber\\
& & {} \sum_{pol, f} \epsilon_{f,\beta} \epsilon_{f, \beta'} = g_{\beta \beta'} - \frac{k_{f, \beta} \ k_{f, \beta'}}{(m_g^f)^2}.
\end{eqnarray}}

  We recall that for vector fields with nonzero Lagrangian mass there is no gauge freedom anymore. The massive vector field  $A_{\mu}$ only has to fulfil  the condition $\partial^{\mu} A_{\mu} = 0$. Therefore, the sum over the initial and final gluon polarizations is fixed by the expressions in (\ref{equ:Sec3.5}).

  For the case of finite masses and widths of the scattering quarks and gluons, the $gQ$ elastic scattering amplitude has been given in Ref. \cite{Berrehrah:2013mua}, where we have to take into account the spectral functions for the heavy quark and gluon masses at finite temperature and chemical potential, the coupling  (cf. Fig.\ref{fig:DQPM-alpha}) at finite $T$ and $\mu_q$ and the quark and gluon propagators for the case of massive vector gluons with finite lifetime $G_F^{\mu \nu} (q, m_g)$ and for the case of massive fermions with finite life time $S_F (p, m_q)$:

{\setlength\arraycolsep{-5pt}
\begin{eqnarray}
\label{equ:Sec3.6}
& & G_F^{t, \mu \nu} (q) = - i \frac{g^{\mu \nu} - q^{\mu} q^{\nu}/m_g^2}{t - m_g^2 + i 2 \gamma_g (p_0^i - p_0^f)}, \hspace{0.3cm} S_F^{u} (p) = \! \frac{\slashed{p} + M_Q}{u - M_Q^2 + i 2 \gamma_Q (p_0^i - k_0^f)}, \hspace{0.3cm} S_F^{s} (p) = \! \frac{\slashed{p} + M_Q}{s - M_Q^2 + i 2 \gamma_Q (p_0^i + p_0^f)},
%\nonumber\\
%& & {}
\end{eqnarray}}

 where $m_g$, $\gamma_g$ ($M_Q$, $\gamma_Q$) are the mass and width of the gluon or the heavy quark at finite temperature and chemical potential (cf. Sec.\ref{DQP-finiteTmu}). We note that the heat bath breaks the Lorentz covariance. Accordingly the energies $p_0^i$ and $ p_0^f$ denote quantities in the rest frame of the heat bath.

%------------------------------------------------------------------------------------------------------------------------------------------------------------------------------
\subsection[Results]{Results for elastic scatterings} %------------------------------------------------------------------------------------------------------------------------------------------------------------------------------

\begin{figure}[h!]
       \begin{center}
       \includegraphics[width=8.8cm, height=8.5cm]{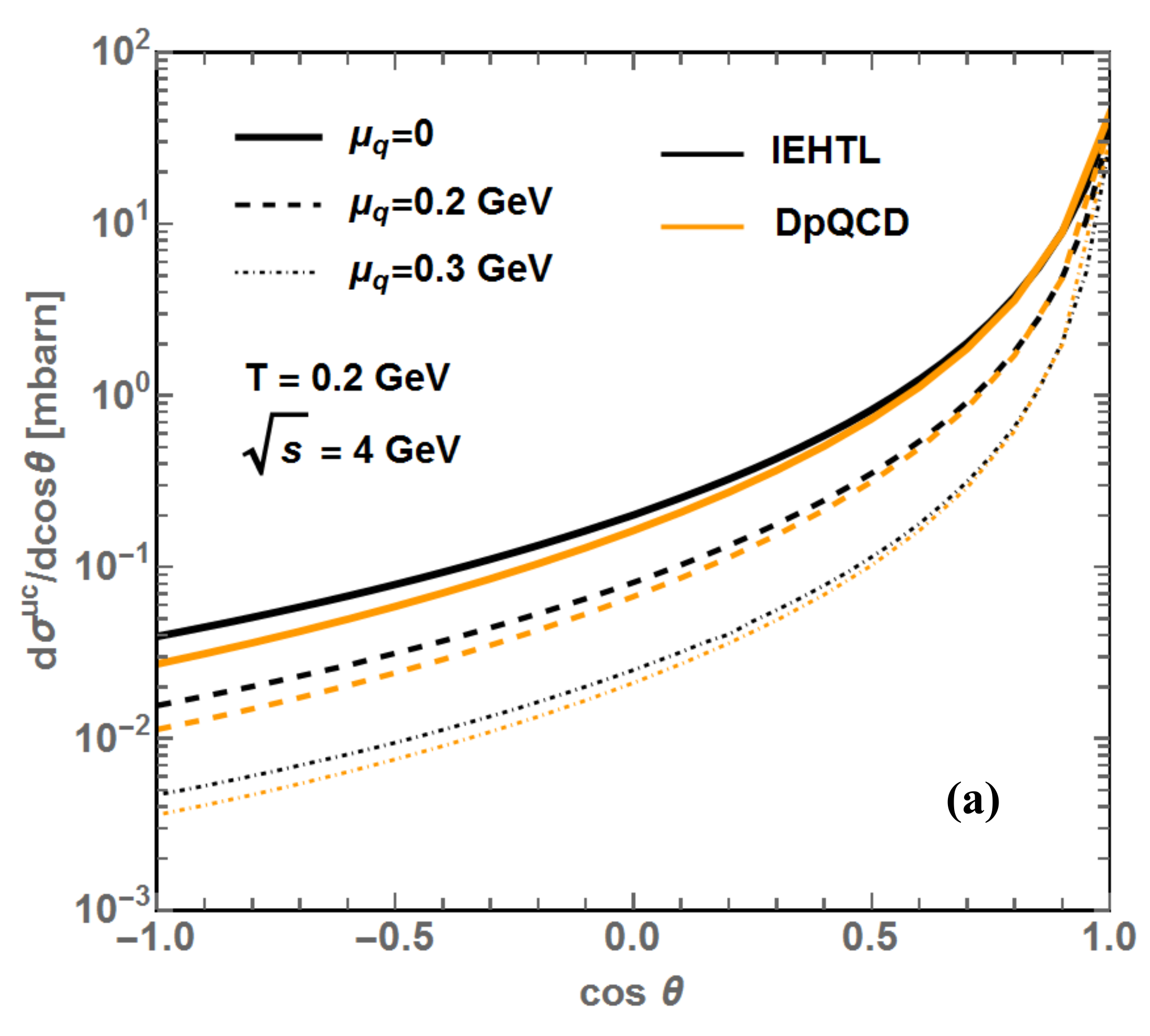}
       %\\ \vspace*{0.3cm}
       \includegraphics[width=8.8cm, height=8.5cm]{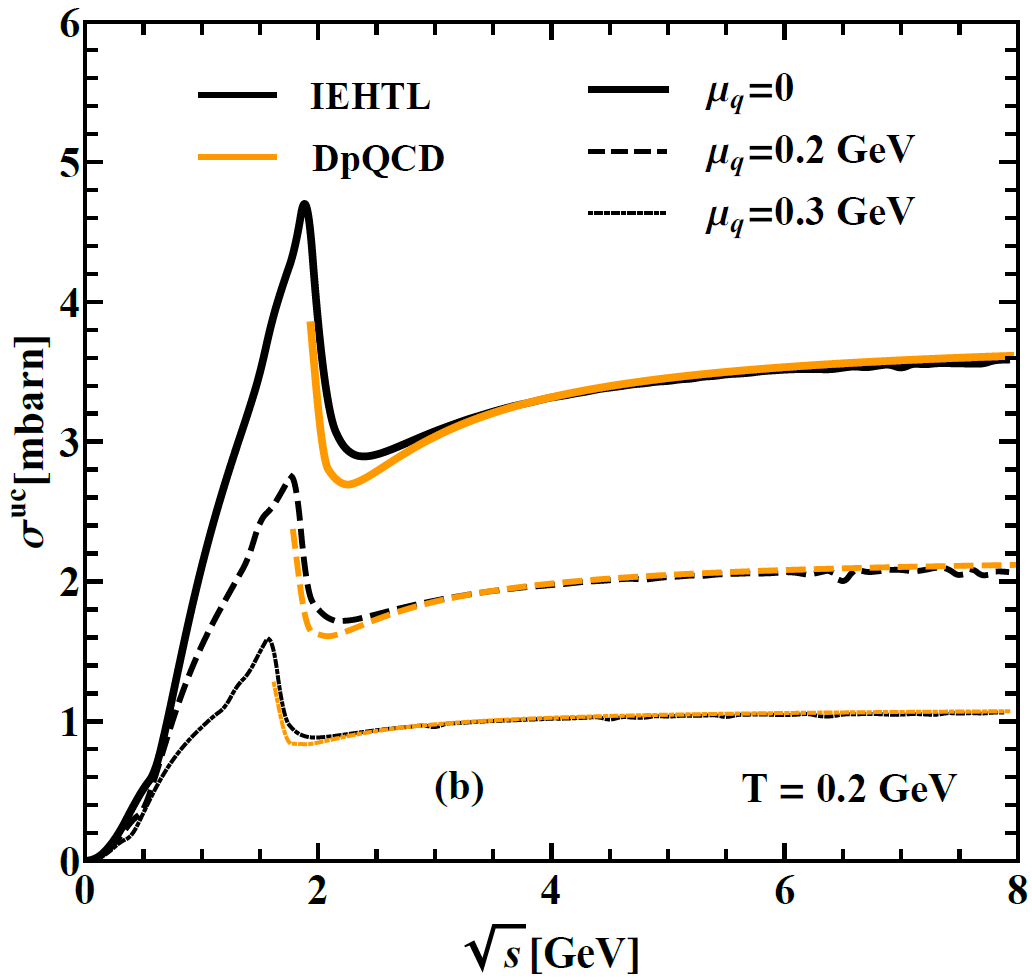}
       \end{center}
\vspace*{-0.3cm}
\caption{\emph{(Color online) Differential (a) and total (b) elastic cross section for $u c \rightarrow u c$ elastic scattering for off-shell (black lines) and on-shell partons (orange lines) at three different values of the quark chemical potential $\mu_q$ (see legend) at $T = 0.2$ GeV. We consider the DQPM pole masses for the on-shell partons and the DQPM spectral functions for the off-shell degrees of freedom.} \label{fig:dSigdthetaSigqQDpQCD-IEHTL}}
\end{figure}

   We consider first the scattering of a (high momentum) heavy quark with a light quark or a gluon in a QGP at temperature $T$ = 0.2 GeV with invariant energy $\sqrt{s} = 4$ GeV for different quark chemical potential $\mu_q = 0, 0.2, 0.3$ GeV. Fig. \ref{fig:dSigdthetaSigqQDpQCD-IEHTL}-(a) presents the off-shell differential cross section $d \sigma\!/\!d \cos \theta$ (black lines) in comparison to the on-shell cross section (orange lines) of the $u c$ elastic scattering. The importance of finite width corrections in the $u c$ scattering processes is illustrated by comparing the two differential cross sections. For the energy of $\sqrt{s} = 4$ GeV one observes a deviation of the off-shell results compared to the on-shell ones only for large scattering angles. However, according to the small differences between the differential on- and off-shell cross sections one can conclude that the total on-shell cross sections do not change on a relevant scale when introducing off-shell masses. This is particularly important since the width of the heavy quark has been taken as an upper limit (cf. Section \ref{DQP-finiteTmu}). Figure \ref{fig:dSigdthetaSigqQDpQCD-IEHTL}-(a) shows also the influence of a finite chemical potential on the heavy quark scattering. The increase of $\mu_q$ leads to a decrease of $d \sigma\!/\!d \cos \theta$ and consequently to a  decrease of the total cross section as illustrated in Fig.\ref{fig:dSigdthetaSigqQDpQCD-IEHTL}-(b). Despite the lower value of the IR regulator at higher values of $\mu_q$ (decrease of the gluon mass at finite $\mu_q$) the smaller values of the running coupling at finite $\mu_q$ explain the decrease of $d \sigma\!/\!d \cos \theta$ at finite and large values of $\mu_q$.

   The total elastic cross section of a $c$-quark, which traverses a plasma at temperatures $T=0.2$ GeV, as calculated in the DpQCD and IEHTL approaches, is shown in Fig. \ref{fig:dSigdthetaSigqQDpQCD-IEHTL}-(b) as a function of $\sqrt{s}$ for different values of $\mu_q$. Apart for energies close to the threshold the cross sections show a rather smooth dependence on the invariant energy $\sqrt{s}$, however, differ substantially in magnitude with temperature and chemical potential. Fig. \ref{fig:dSigdthetaSigqQDpQCD-IEHTL}-(b) demonstrates also that, independently on $\mu_q$, the off-shell mass distributions only have a sizeable impact at the threshold given by the pole masses for $uc$ scattering. This is due to the moderate parton widths considered in the DQPM model. At energies below the on-shell threshold the off-shell cross section increases with $\sqrt{s}$ because more and more masses can contribute. Whereas the on-shell cross section diverges at the threshold the off-shell cross section shows a maximum at the on-shell threshold and decreases then due to the decrease of the on-shell cross section.

   The conclusions drawn before for the study of $uc$ elastic scattering are valid also for  $gc$ scattering, however, with cross sections for  $gQ$ elastic scattering that are larger than the cross sections for $uc$ scattering by roughly a factor of $9/4$ which is ratio of the different color Casimir operators (squared). This is also related to the fact that the scattering of heavy quarks with gluons proceeds via $t$, $s$ and $u$ channels, whereas one has only the $t$ channel for $uc$ elastic scattering.

\begin{figure}[h!]
       \begin{center}
       \includegraphics[width=8.9cm, height=8.6cm]{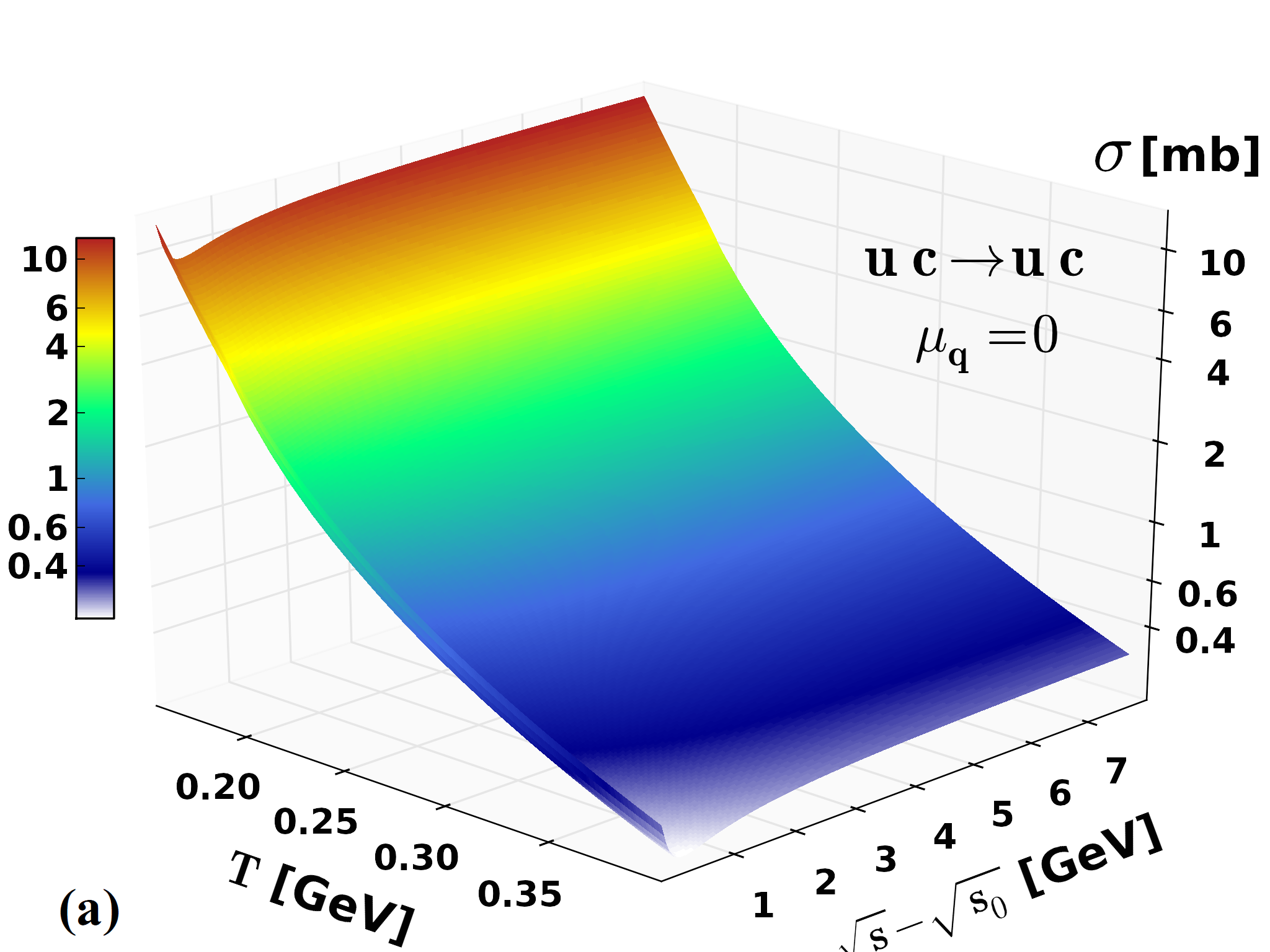}
       \includegraphics[width=8.9cm, height=8.6cm]{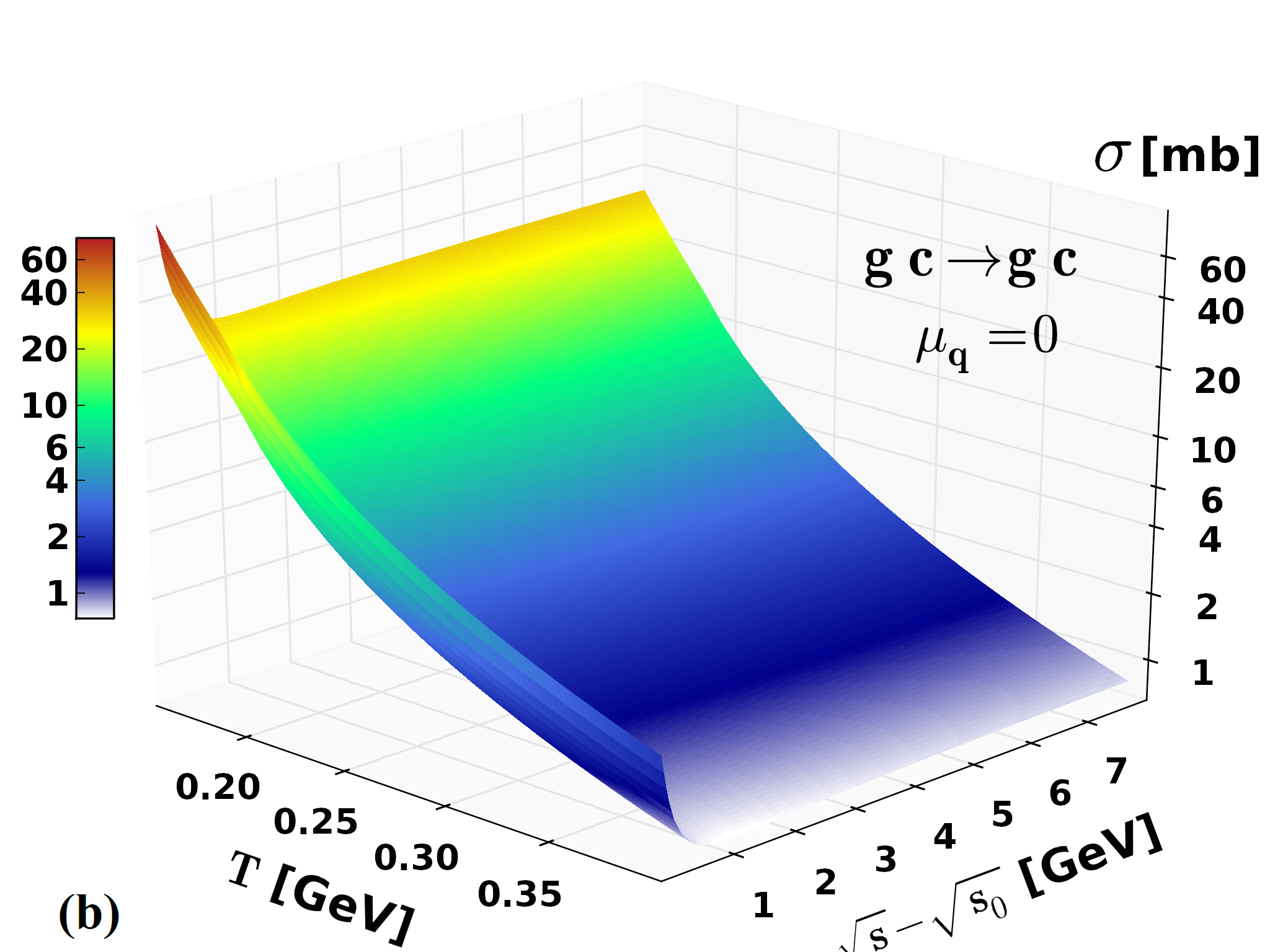}
       \end{center}
\vspace*{-0.2cm}
\caption{\emph{(Color online) Elastic cross section of $u c \rightarrow u c$ (a) and $g c \rightarrow g c$ (b) scattering as a function of the temperature $T$ and the invariant energy above threshold $\sqrt{s} - \sqrt{s_0} $, where $\sqrt{s_0}$ is the threshold energy, for on-shell partons as described by the DpQCD approach at $\mu_q = 0$.} \label{fig:Sigugc-DpQCD-IEHTL_vsTS}}
\end{figure}

   Figs.\ref{fig:Sigugc-DpQCD-IEHTL_vsTS}-(a) and (b) show explicitly the temperature and $\sqrt{s}$ dependences of the $uc$ and $gc$ elastic cross sections at $\mu_q = 0$, as described in the DpQCD approach. We deduce that an increasing medium temperature $T$ leads to an increase of the thermal gluon mass (infrared regulator) and hence to a decrease of the DpQCD $uc$ and $gc$ elastic cross sections. The effective gluon mass is roughly proportional to $T$ for temperatures above $0.2$ GeV. The large enhancement of the total cross section for temperatures close to $T_c (\mu_q)$ can be traced back to the infrared enhanced coupling in the DpQCD/IEHTL models.

\begin{figure}[h!]
       \begin{center}
       \includegraphics[width=8.9cm, height=8.6cm]{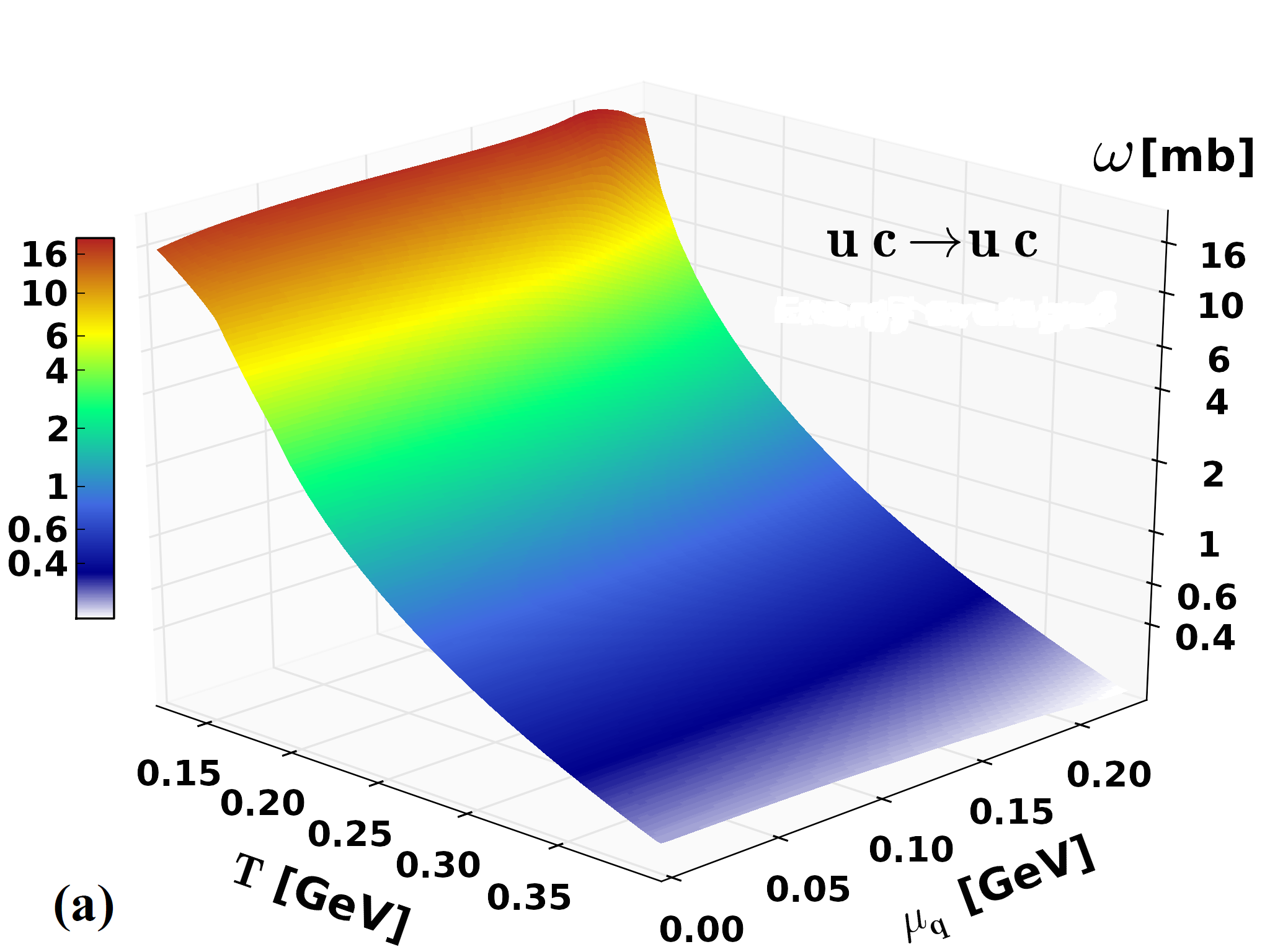}
       \includegraphics[width=8.9cm, height=8.6cm]{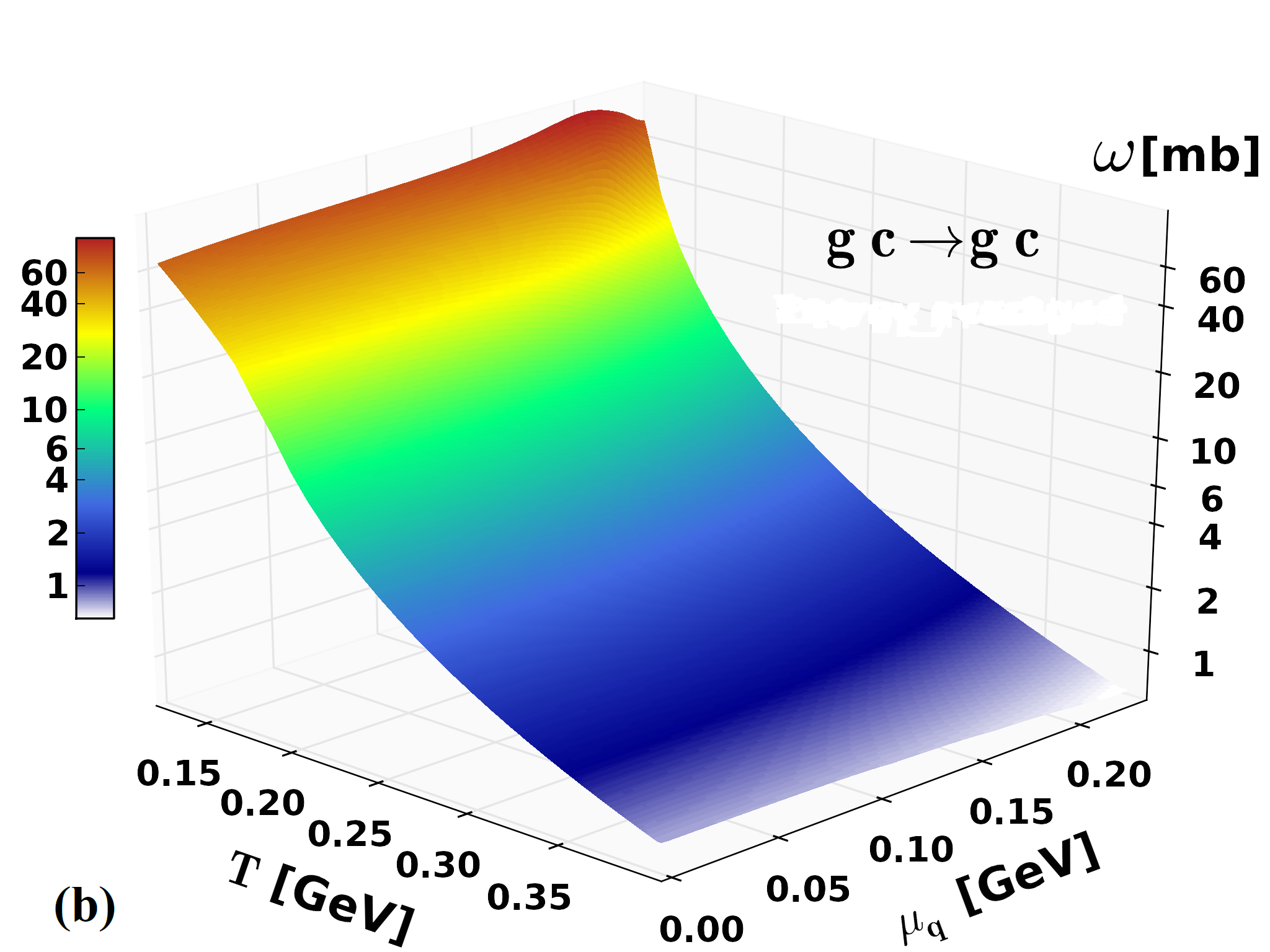}
       \end{center}
\vspace{-0.5cm}
\caption{\emph{(Color online) Thermal transition rate $\omega$ of $u c \rightarrow u c$ (a) and $g c \rightarrow g c$ (b) as a function of the temperature $T$ and quark chemical potential $\mu_q$ for on-shell partons as described by the DpQCD approach}. \label{fig:MeansSigugc-DpQCD-IEHTL_vsTmu}}
\end{figure}

    In order to quantify the effect of a finite $\mu_q$ on the heavy quark scattering, we show in Figs. \ref{fig:MeansSigugc-DpQCD-IEHTL_vsTmu}-(a) and (b) the $uc$ and $gc$ thermal transition rate $\omega$ (defined by Eq. (3.8) in Ref. \cite{Berrehrah:2014kba}) as a function of $T$ and $\mu_q$. We provide the results only for DpQCD because there is just a small difference between the on-shell DpQCD and the off-shell IEHLT approaches (cf. Fig.\ref{fig:dSigdthetaSigqQDpQCD-IEHTL}).

   There are two different profiles in the $(T,\mu_q)$ dependencies of the thermal transition rate $\omega$. For temperatures larger than $T_c (\mu_q = 0) = 0.158$ GeV, a small effect of a finite $\mu_q$, leading to a decrease of $\omega$, is noticed. For temperatures smaller than $T_c (\mu_q=0)$, an increase of $\omega$ appears when $\mu_q$ increases. The last effect is due to the increase of the coupling $\alpha_s$ at these temperatures. More precisely, one has $\alpha_s (T < T_c(\mu_q=0), \mu_q) > \alpha_s (T = T_c(\mu_q=0), \mu_q)$. Besides the transition amplitudes, one should also consider the effect of the statistical weights $f_{uc}^{FD}$ (Fermi-Dirac distribution function) at finite $(T, \mu_q)$ in the evaluation of $\omega$; $f_{uc}^{FD}$ increases for large values of $\mu_q$ leading to an extra contribution to the increase of the thermal transition rate $\omega$ for $T < T_c (\mu_q = 0)$, but is not enough to counterbalance the decrease of the total cross section for $T > T_c (\mu_q = 0)$.

  The thermal transition rate can be parametrized by a power law in $T$ for each value of $\mu_q$, i.e. $\sim \displaystyle T^{-\beta}$ for $T > T_c (\mu_q = 0)$. In fact, one can find that $\displaystyle \beta^{T_c(\mu_q=0)<T<1.2 T_c(\mu_q)} \sim 4, \beta^{T>1.2 T_c (\mu_q)} \sim 2$. These different power laws in $T$ will have a sizeable effect on the transport coefficients to be evaluated in the following sections. On the other hand, having almost the same power laws at finite $\mu_q$ as compared to $\mu_q = 0$ (for temperatures larger than $T_c(\mu_q=0)$) leads to some scaling effects in the transport coefficients at these temperatures.

\begin{figure}[h!]
       \begin{center}
       \includegraphics[width=8.7cm, height=8.2cm]{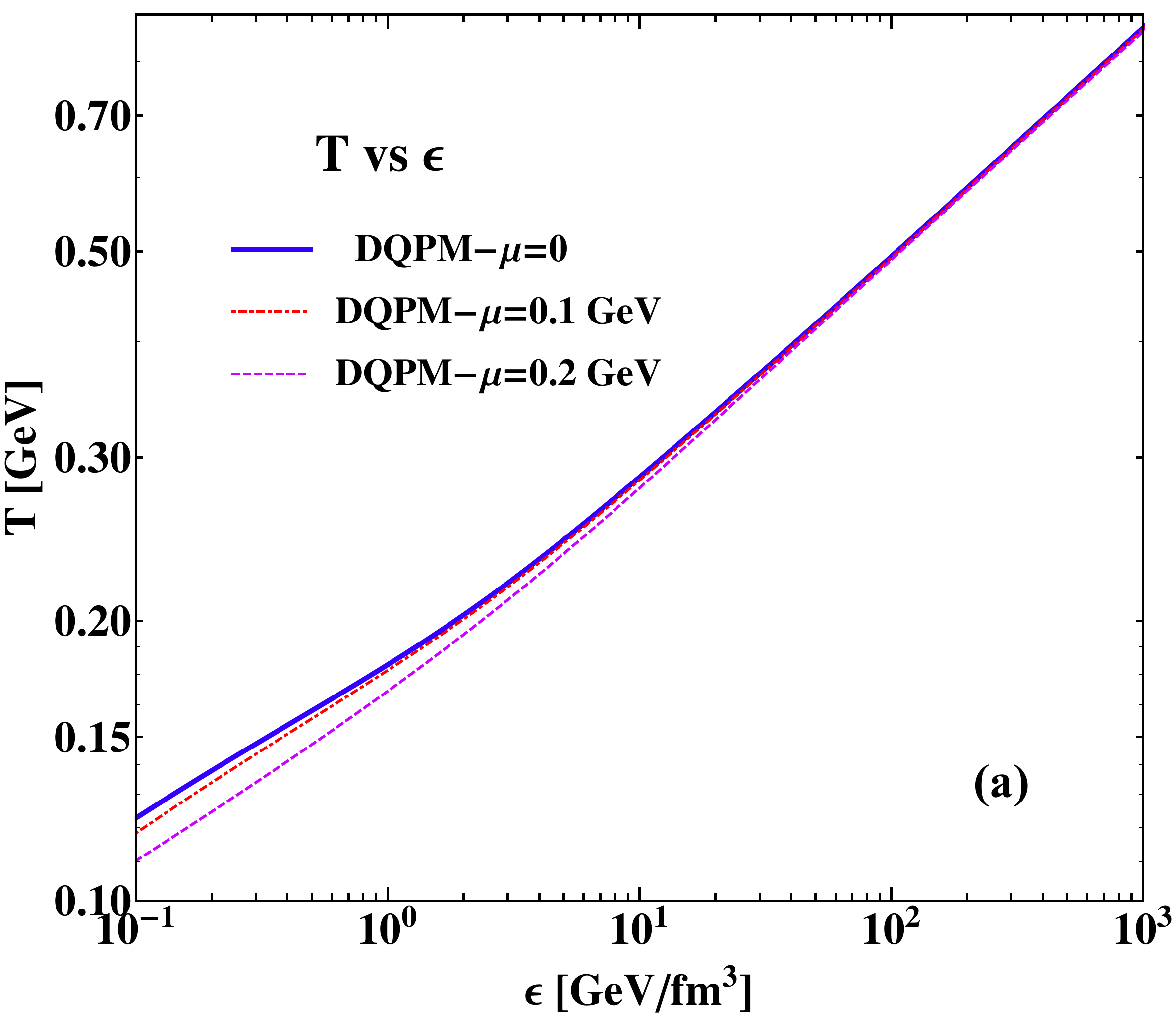}
       \includegraphics[width=8.7cm, height=8.2cm]{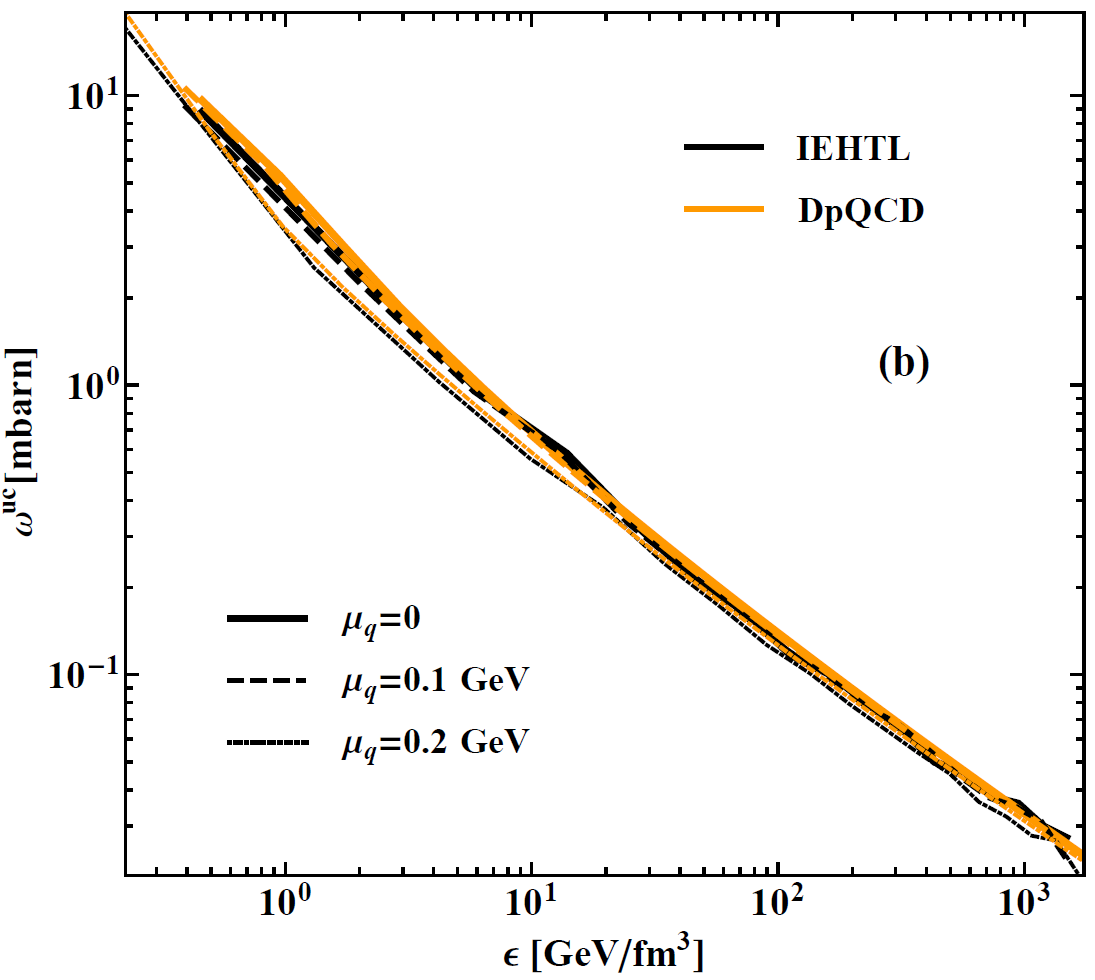}
       \end{center}
\vspace*{-0.4cm}
\caption{\emph{(Color online) Temperature $T$ as a function of the energy density $\epsilon$ from the DQPM for different values of the quark chemical potential $\mu_q$ (a); Thermal transtion rate for $u c \rightarrow u c$ scattering for off-shell (IEHTL) and on-shell (DpQCD) partons as a function of the energy density $\epsilon$ (b).} \label{fig:TvsEpsMu-Sigmauc}}
\end{figure}

  Additionally, it is worth to evaluate the $uc$ and $g c$ elastic scattering cross section as a function of the energy density $\epsilon$ available in the heavy-ion collisions since out-off equilibrium a temperature is ill defined while $\epsilon$ can be well calculated in the local rest frame (e.g. in PHSD). Here the energy density $\epsilon$ for a given temperature $T$ and quark chemical potential $\mu_q$ is obtained by using the inverted DQPM equation of state which gives the temperature as a function of the energy density $\epsilon$ for a given quark chemical potential. The temperature as a function of $\epsilon$ for $\mu_q = 0, 0.1, 0.2$ GeV is shown in Fig. \ref{fig:TvsEpsMu-Sigmauc}-(a). Note that the DQPM model describes the QCD energy density at temperatures even as low as $T \sim T_c (\mu_q)$.

  Transport theoretical simulations have shown that the local energy densities achieved in the course of heavy-ion collisions at FAIR energies reach at most 3 GeV fm$^{-3}$, at SPS and RHIC energies up to 30 GeV fm$^{-3}$ and up to 300 GeV fm$^{-3}$ at LHC. Therefore one observes that the $uc$ elastic cross section, displayed in Fig. \ref{fig:TvsEpsMu-Sigmauc}-(b) at the energy densities of interest is $\in$ [0.08-8] mb from LHC to FAIR energies, following the DpQCD/IEHTL approaches, with typical values of 5 mb at the phase transition line.
  %------------------------------------------------------------------------------------------------------------------------------------------------------------------------------
\section{Heavy quark interaction rates in a medium at finite $T$ and $\mu_q$}
\label{HQRates}
%------------------------------------------------------------------------------------------------------------------------------------------------------------------------------
\vskip -2mm

  Using the elastic cross section for $q (\bar{q})Q$ and $g Q$ collisions, for on- and off-shell partons -- as calculated in Sec.\ref{qgQProcess} --  we evaluate the interaction rate  of a heavy quark with momentum $\bsp{p}$ and energy $E$ propagating through a QGP in thermal and chemical equilibrium at a given temperature $T$ and quark chemical potential $\mu_q$. The quarks/antiquarks of the plasma are described by a Fermi-Dirac distribution $f_{q,\bar{q}} (\bsq{q}) = \frac{1}{e^{(E_q \mp \mu_q)/ T} + 1}$ whereas the gluons follow a Bose-Einstein distribution $f_{g} (\bsq{q}) = \frac{1}{e^{E_g/ T} - 1}$.

   For on-shell particles (DpQCD model) and in the reference system in which the heavy quark has the velocity $\bsbt{\beta} = \bs{p}/E$ the (on-shell)  interaction rate $R^{\textrm{on}} (\bs{p}) = \frac{d N_{coll}^{2 \rightarrow 2}}{dt}$ for $2 \rightarrow 2$ collisions is given by \cite{Berrehrah:2014kba}

{\setlength\arraycolsep{0pt}
\begin{eqnarray}
\label{equ:Sec4.1}
R^{\textrm{on}} (\bs{p}, T, \mu_q) = \displaystyle \sum_{q,\bar{q},g} \frac{M_Q}{16 (2 \pi)^4 E} \int \frac{q^3 m_0^{\textrm{on}} (s) f_r (\bsq{q})}{s \ E_q} \ d q,
\end{eqnarray}}

  where $\displaystyle \sum_{q,\bar{q},g}$ denotes the sum over the light quarks/antiquarks and gluons of the medium. In Eq. (\ref{equ:Sec4.1})  $f_r (\bsq{q})$ is the invariant distribution of the plasma constituents in the rest frame of the heavy quark, given for the quark/antiquark by:

{\setlength\arraycolsep{0pt}
\begin{eqnarray}
\label{equ:Sec4.2}
\int d\Omega\ f_r (\bsq{q})= 2\pi \int d cos\theta_r \frac{1}{e^{(u^0 E_q - u \ q \cos \theta_r \mp \mu_q)/T} + 1},
\end{eqnarray}}

 with $u\equiv (u^0,\bsu{u}) = \frac{1}{M_Q}(E,-\bs{p})$ being the fluid 4-velocity measured in the heavy-quark rest frame, while $\theta_r$ is the angle between $\bsq{q}$ and $\bsu{u}$. $m_0^{\textrm{on}} (s)$ in (\ref{equ:Sec4.1}) is related to the transition amplitude $|\mathcal{M}_{2,2}|^2$ of the collision $q(\bar{q},g) Q \rightarrow q(\bar{q},g) Q$ by

{\setlength\arraycolsep{0pt}
\begin{eqnarray}
\label{equ:Sec4.3}
m_0^{\textrm{on}} (s) = \frac{1}{2 p_{cm}^2 (s)} \!\! \int_{- 4 p_{cm}^2}^0 \  \frac{1}{g_Q g_p} \sum_{i,j} \sum_{k,l} |\mathcal{M}_{2,2} (s, t; i, j | k, l)|^2 \ d t,
\end{eqnarray}}

 with $p_{cm} = (q \ M_Q)/\sqrt{s}$ denoting the momentum of the scattering partners in the c.m. frame and $g_Q$ ($g_p$)  the degeneracy factor of the heavy quark (parton).

  For off-shell partons (IEHTL model) the elastic interaction rate is obtained by replacing  \begin{equation} \int \frac{d^3 p}{(2 \pi)^3} \frac{1}{2 E} \ \rightarrow \  \int \frac{d^4 p}{(2 \pi)^4} \rho (p) \Theta (p_0) \label{equ:OnOff} \end{equation} with $\rho (p)$ denoting the spectral function which can be specific for each particle species. For the Breit-Wigner-$m$ form of the spectral function, the explicit extension of Eq.(\ref{equ:Sec4.1}) for the off-shell case is given by

{\setlength\arraycolsep{0pt}
\begin{eqnarray}
\label{equ:Sec4.4}
R^{\textrm{off}} (\bs{p}, T, \mu_q) = \sum_{q,\bar{q},g} \Pi\displaystyle_{i \in{p,q,p',q'}} \ \int m_i d m_i \rho_i^{BW} (m_i) \frac{m_p}{16 (2 \pi)^4 E} \int \frac{q^3 m_0^{\textrm{off}} (s) f_r (\bsq{q})}{s \ E_q} \ d q,
\end{eqnarray}}
 where
{\setlength\arraycolsep{0pt}
\begin{eqnarray}
\label{equ:Sec4.5}
& & m_0^{\textrm{off}} (s) = \frac{1}{2 p^i p^f} \int_{t_{min}}^{t_{max}} \frac{1}{g_p g_Q} \sum_{i,j} \sum_{k,l} |\mathcal{M}_{2,2}^{\textrm{off}} (s, t; i, j | k, l)|^2 \ d t,
%\nonumber\\
%& & {}
\end{eqnarray}}

  with $\sum |\mathcal{M}_{2,2}^{\textrm{off}} (p, q; i, j |p', q'; k, l)|^2 (s,t)$ being the off-shell transition amplitude defined in the IEHTL approach (cf. Ref.\cite{Berrehrah:2013mua}); $p^i$ ($p^f$) is the initial (final) heavy-quark momentum and the integration boundaries $t_{min}$ and $t_{max}$ are given by:

{\setlength\arraycolsep{0pt}
\begin{eqnarray}
\label{equ:Sec4.6}
& & t_{min}^{max} = - \frac{s}{2} \Biggl(1 - (\beta_1 + \beta_2 + \beta_3 + \beta_4) + (\beta_1 -  \beta_2)(\beta_3 - \beta_4) \pm \sqrt{(1 - \beta_1 - \beta_2)^2 - 4 \beta_1 \beta_2} \sqrt{(1 - \beta_3 - \beta_4)^2 - 4 \beta_3 \beta_4} \Biggr),
%\nonumber\\
%& & {} t_{min}^{max} = - \frac{s}{2} (C_1 \pm C_2)
%\nonumber\\
%& & {} \textrm{where:} \ C_1 = 1 - (\beta_1 + \beta_2 + \beta_3 + \beta_4) + (\beta_1 -  \beta_2)(\beta_3 - \beta_4)
%\nonumber\\
%& & {} C_2 = \sqrt{(1 - \beta_1 - \beta_2)^2 - 4 \beta_1 \beta_2} \sqrt{(1 - \beta_3 - \beta_4)^2 - 4 \beta_3 \beta_4}
\nonumber\\
& & {} \textrm{with:} \ \beta_1 = (m_q^i)^2/s, \beta_2 = (M_Q^i)^2/s, \beta_3 = (m_q^f)^2/s, \beta_4 = (M_Q^f)^2/s.
\end{eqnarray}}

\begin{figure}[h!] %tbh!
\begin{center}
   \begin{minipage}[c]{.48\linewidth}
     \includegraphics[height=8.6cm, width=8.8cm]{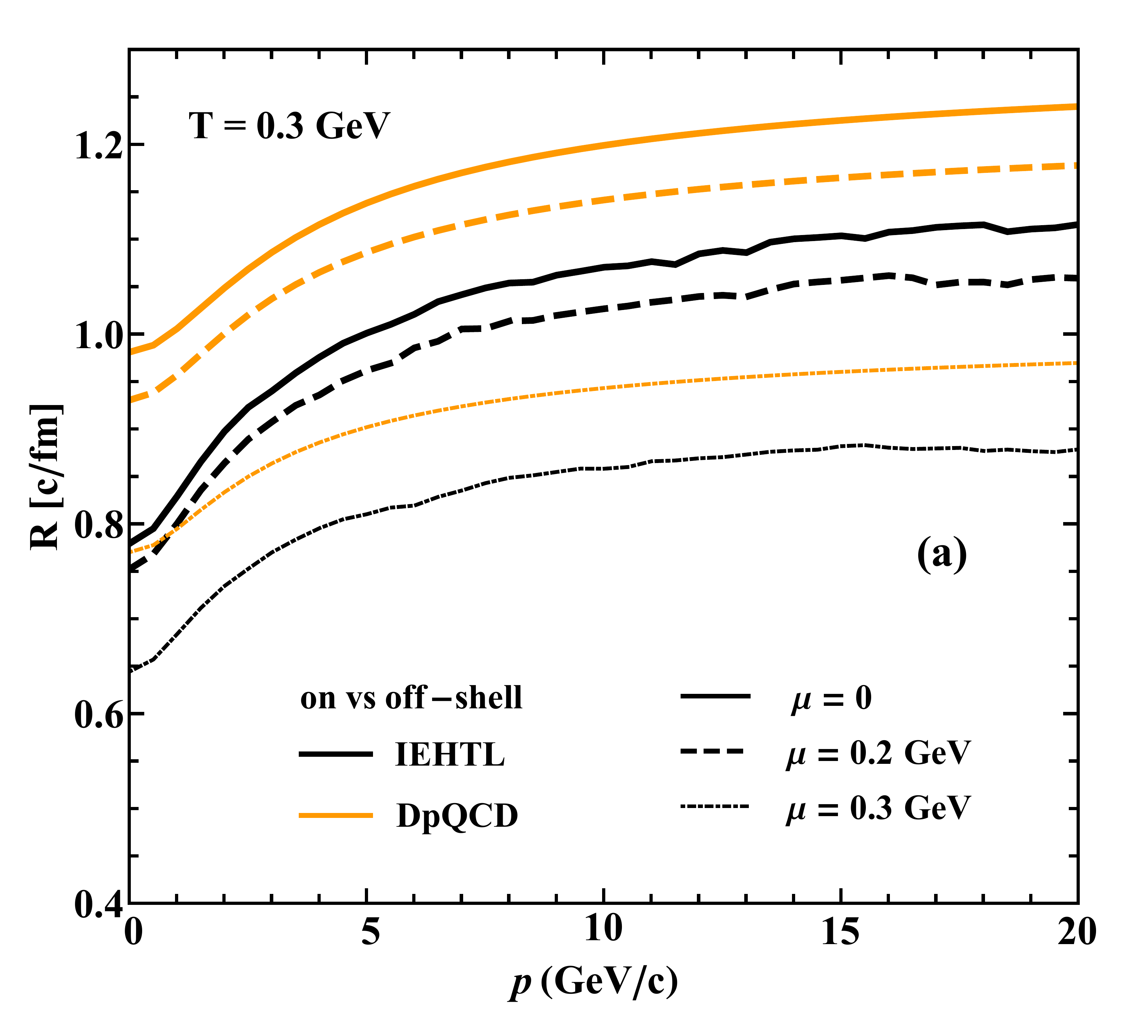}
   \end{minipage} \hfill
   \begin{minipage}[c]{.48\linewidth}
     \includegraphics[height=8.3cm, width=8.8cm]{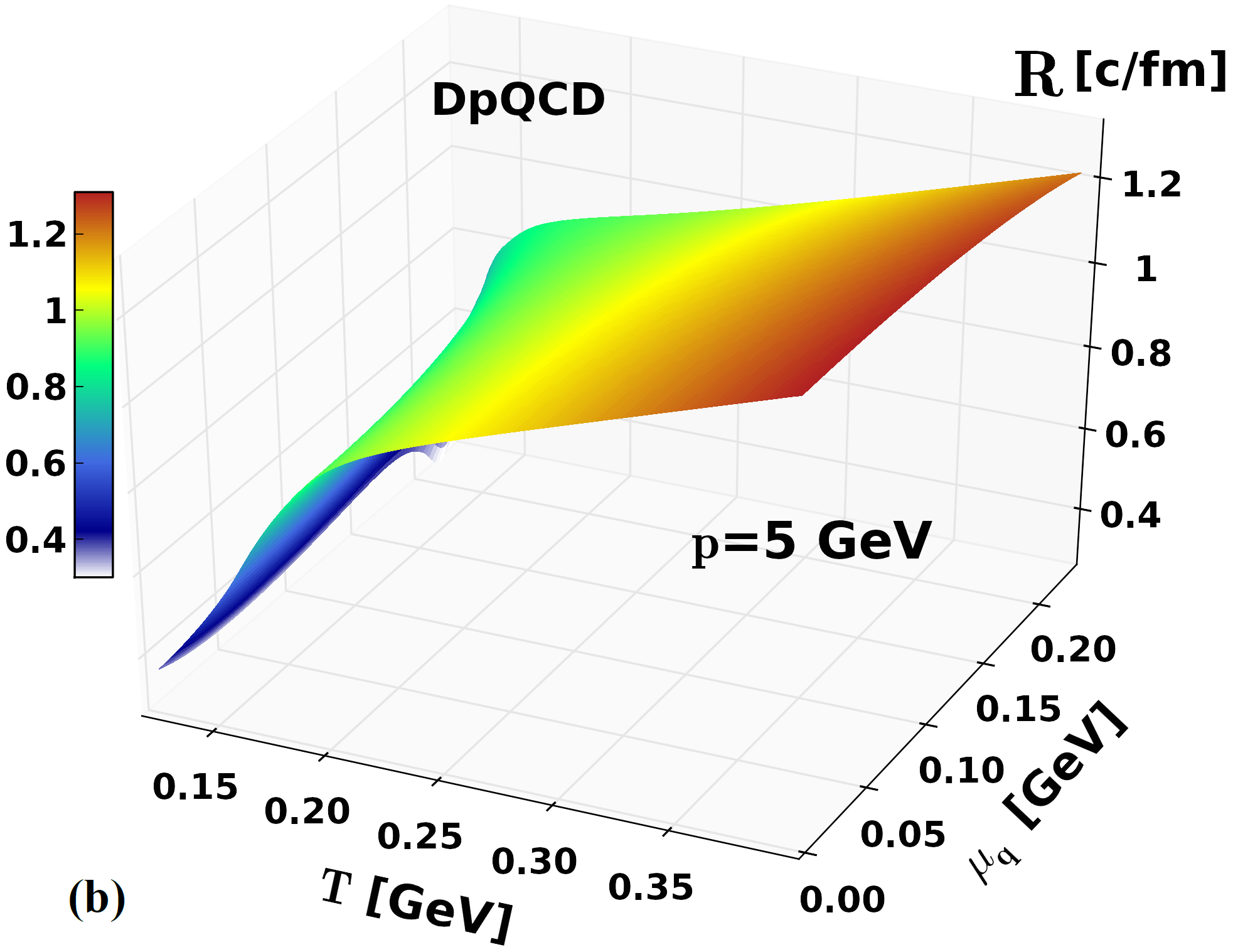}
\end{minipage}
\vspace{-0.1cm}
\caption{\emph{(Color online)  The total elastic interaction rate $R$ of c-quarks in the plasma rest frame in the IEHTL and DpQCD models as a function of the heavy-quark  momentum $p$  for three different values of the quark chemical potential $\mu_q = 0, 0.2, 0.3$ GeV at $T = 0.3$ GeV (a). (b) Contour plot of the total elastic interaction rate $R$ within DpQCD as a function of $T$ and $\mu_q$.}}
\label{fig:cRatevspQ}
\end{center}
\end{figure}

 The total interaction rate (\ref{equ:Sec4.4}) of the off-shell approach (IEHTL) in the plasma rest system is compared to that of the on-shell calculations (DpQCD) (\ref{equ:Sec4.1}) in Fig. \ref{fig:cRatevspQ}-(a) as a function of the momentum of the heavy quark $p$ for different values of the quark chemical potential $\mu_q$ at $T$ = 0.2 GeV. We assume here a Breit-Wigner spectral function and a Boltzmann-J\"utner distribution for both, the light quarks/antiquarks and the gluons. Our results are rather independent of the choice of the spectral function or by replacing the Boltzmann-J\"utner distribution by a Fermi/Bose distribution. In Fig. \ref{fig:cRatevspQ}-(a) the black lines refer to IEHTL results and the orange lines to DpQCD results. We see from Fig. \ref{fig:cRatevspQ}-(a) that the finite width of the spectral function (with the DQPM width) decreases the interaction rate of heavy quarks with the medium on the order of 20\%. This modification is rather independent of the heavy-quark momentum, temperature and quark chemical potential of the plasma. The difference between the DpQCD and IEHTL rates is related on one side to the propagator, which contains an additional imaginary part proportional to the gluon width in the IEHTL model, and on the other side to the energy asymmetric contribution of the Breit-Wigner spectral function in IEHTL.

  For fixed temperature $T = 0.3$ GeV the variation of the quark chemical potential $\mu_q$ from $\mu_q = 0$ leads to a decrease of the rate (cf. Fig.\ref{fig:cRatevspQ} (a)). However, e.g.  for the temperature $T = 0.2$ GeV an increasing $\mu_q$ leads to either an increase of the rate (for $\mu_q$ = 0.2 GeV) or to the decrease of $R$ (for $\mu_q$ = 0.3 GeV). Therefore, the dependence of $R$ on both the medium temperature and quark chemical potential is not trivial. The dependence of the rates on the medium temperature $T$ and quark chemical potential $\mu_q$ for a heavy-quark with a momentum $p = 5$ GeV is illustrated in Fig. \ref{fig:cRatevspQ}-(b) in DpQCD. As expected, the rate increases for higher temperatures at $\mu_q=0$ because the number of plasma particles becomes larger ($\sim T^3$). Therefore, the increase of the interaction rate with temperature keeping $\mu_q$ small is seen. For a fixed temperature, the variation of $\mu_q$ leads to different profiles in the rates. A decrease of the total rate for high temperatures when $\mu_q$ increases and an opposite trend for small temperatures ($T < T_c(\mu_q=0)$) is observed. The highest values of the rates are reached for small $\mu_q$ and large temperatures $T$

\begin{figure}[h!] %tbh!
\begin{center}
     \includegraphics[height=6.43cm, width=5.88cm]{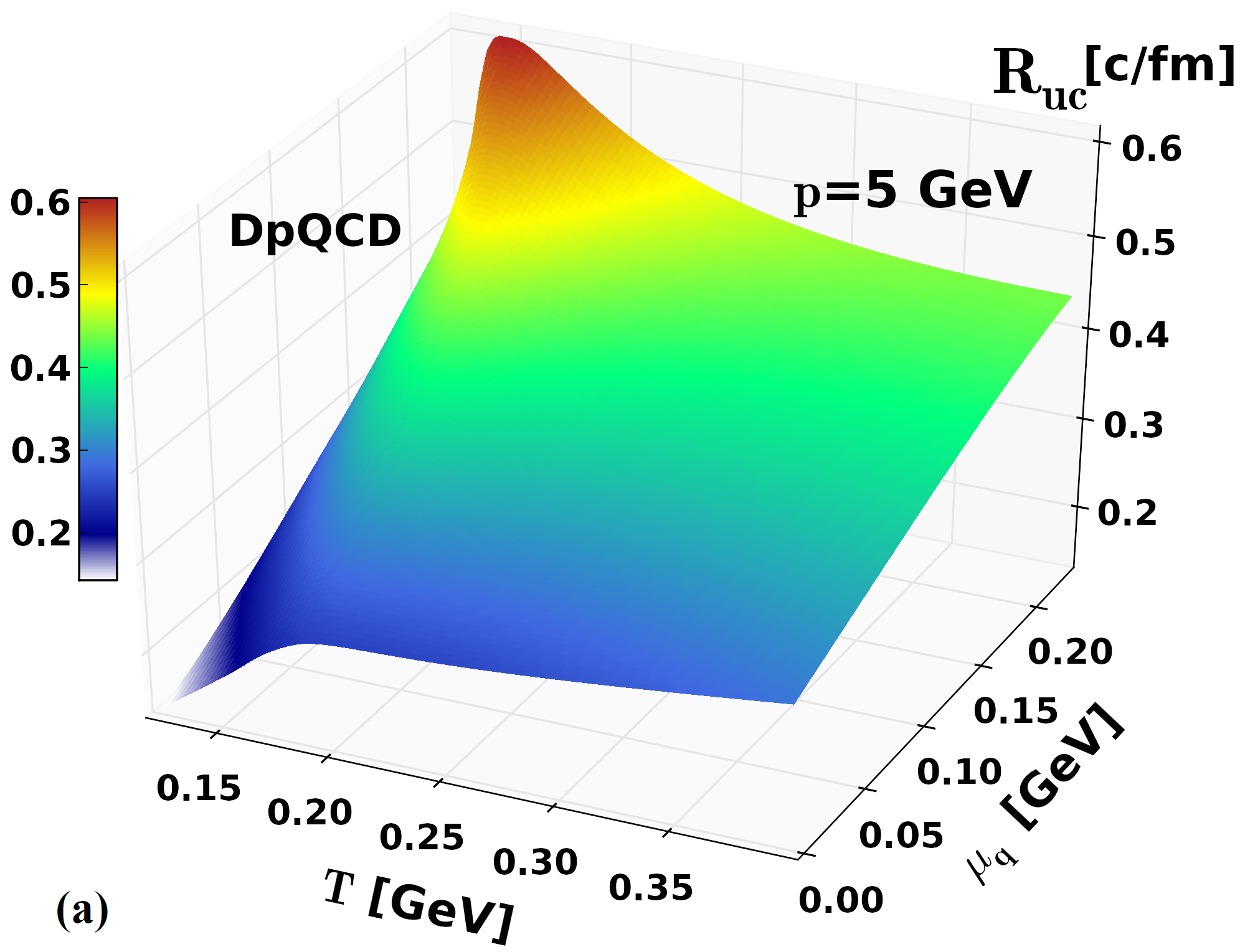}
     \hspace{0.03cm}
     \includegraphics[height=6.43cm, width=5.88cm]{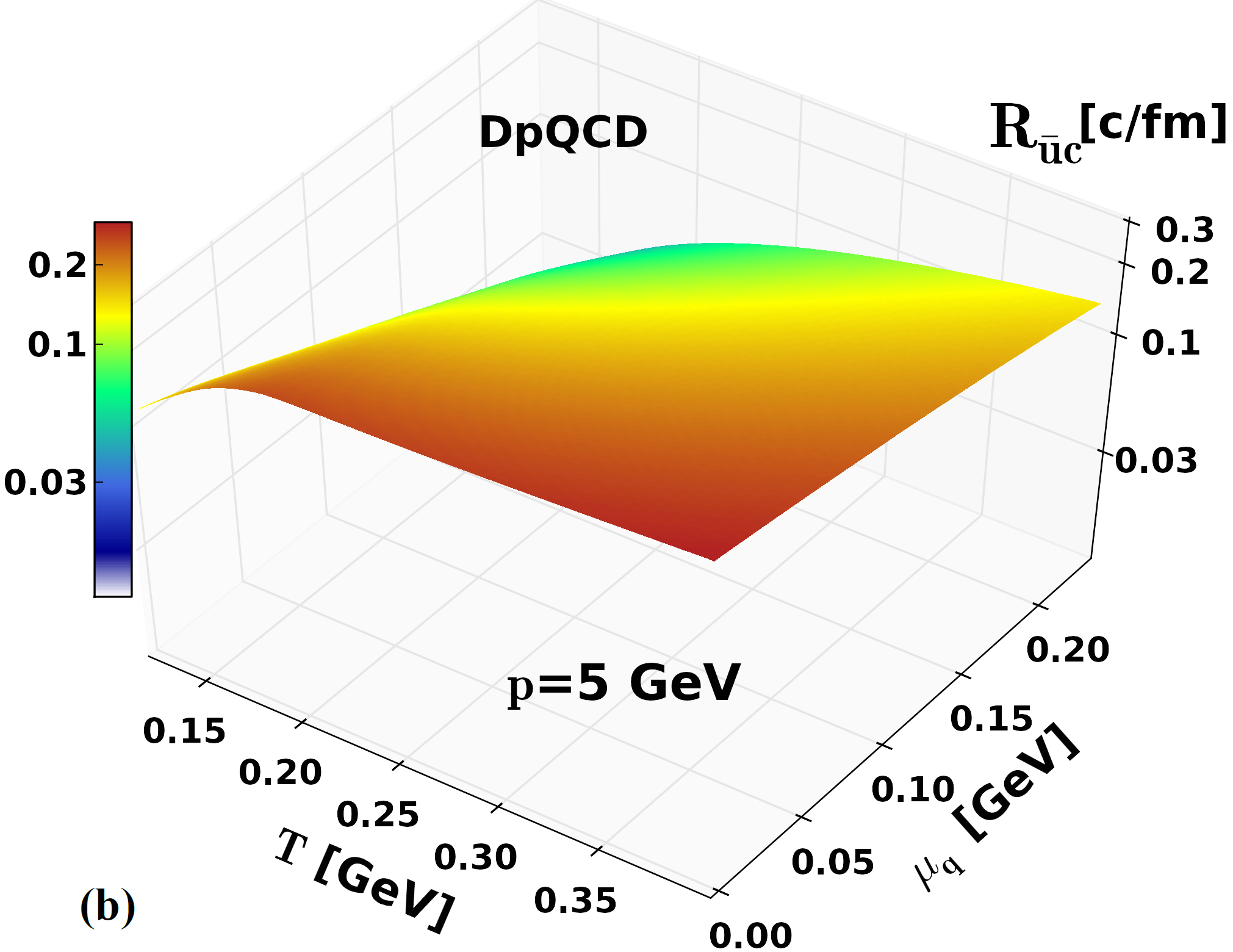}
     \hspace{0.03cm}
     \includegraphics[height=6.43cm, width=5.88cm]{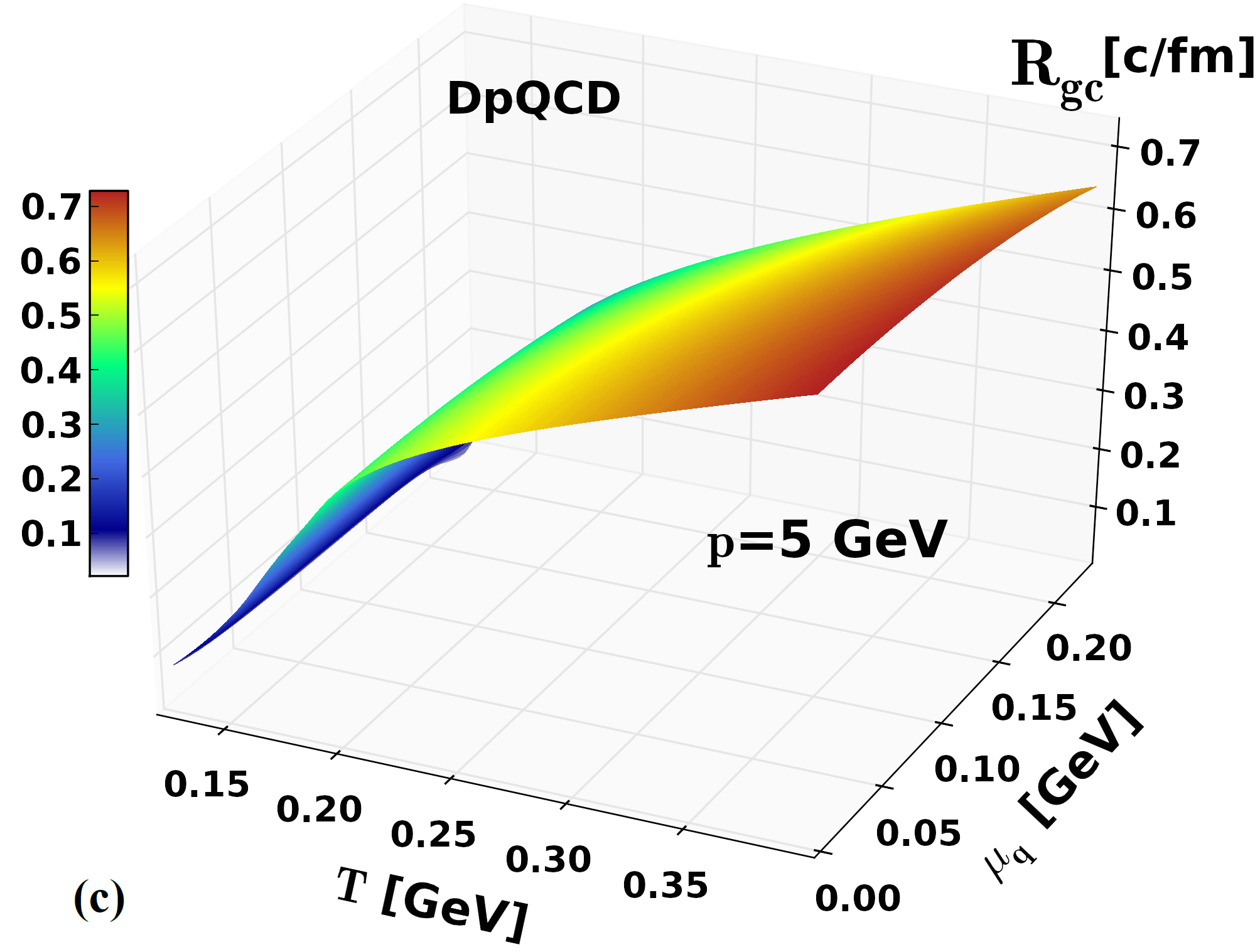}
\vspace{-0.3cm}
\caption{\emph{(Color online) The total elastic interaction rate $R$ of c-quarks in the plasma rest frame as a function of the temperature $T$ and quark chemical potential $\mu_q$ due to the scattering with light quarks (a), light antiquarks (b) and gluons (c). The on-shell heavy-quark momentum in all cases is $p = 5$ GeV.}}
\label{fig:cRatevsTmu}
\end{center}
\end{figure}

  Due to the different abundances of particle species in a medium at finite chemical potential, it is interesting to study the variation of the heavy-quark interaction rates with the quarks/antiquarks and gluons independently. Figs. \ref{fig:cRatevsTmu}-(a), (b) and (c) illustrate the dependence of the heavy-quark collisional rates with quarks, antiquarks and gluons of a medium at finite temperature $T$ and quark chemical potential $\mu_q$  for an intermediate heavy-quark momentum ($p = 5$ GeV).

 For the case of gluons and antiquarks, the interaction rates are increasing with higher temperature for all $\mu_q$. The charm quark interaction rate with light quarks ($R_{uc}$) depends on ($T$,$\mu_q$) in a similar fashion as described in Fig. \ref{fig:cRatevspQ}-(b) for large $\mu_q$ and small temperatures ($T < T_c(\mu_q=0)$). Indeed, for larger values of $\mu_q$ and small temperatures, $R_{uc}$ is much larger than $R_{\bar{u}c}$ and $R_{gc}$, so that the total interaction rates are dominated by $R_{uc}$. On the other hand the $R$ profile is dominated by $R_{gc}$ for small $\mu_q$ and large temperatures. This is easy to interpret: At large $\mu_q$ the number of light quarks is large compared to the number of antiquarks, i.e. the Fermi-Dirac distribution contributes unequally to $R$ for $u$ and $\bar{u}$. On the other hand the gluon number decreases with larger $\mu_q$ since it is correlated with the subdominant light antiquarks, via the $T$ and $\mu_q$ dependencies of the masses.
  %---------------------------------------------------------------------------------------------------------------------------------------------
\section{Heavy quark energy loss in a medium at finite temperature $T$ and chemical potential $\mu_q$}
\label{HQELoss}
%---------------------------------------------------------------------------------------------------------------------------------------------
\vskip -2mm

 The collisional energy loss $dE/dt$ has been formulated by Bjorken and been explicitly calculated in Ref.\cite{Berrehrah:2014kba} for $\mu_q$ = 0. We recall here the expression of $dE/dt$ for on- and off-shell partons. In the framework of the DpQCD model, $dE/dt$ is given in the plasma rest frame by:

{\setlength\arraycolsep{-6pt}
\begin{eqnarray}
\label{equ:Sec5.1}
\frac{d E^{\textrm{on}}}{dt} (p, T, \mu_q) = \frac{M_Q}{4 (2 \pi)^3} \int_0^{\infty} \frac{q^4 m_1^{\textrm{on}} (s)}{s^2 E_q} \ \Biggl[- q f_0 (q) + \frac{p}{E} (M_Q + E_q) f_1 (q) \Biggr] \ d q.
%\nonumber\\
%& & {}
\end{eqnarray}}

  Eq.(\ref{equ:Sec5.1}) is easily extended to the off-shell case with Breit-Wigner spectral functions using (\ref{equ:OnOff}). One finds \cite{Berrehrah:2014kba}
%by replacing \begin{equation}  \frac{1}{2 E}\ \rightarrow  \ \int_0^{\infty} \frac{d \omega}{2 \pi} \rho_p (\omega, \bsp{p}) = \int \frac{d \omega}{2 \pi} \rho_p (\omega, \bsp{p}) \theta (\omega). \end{equation}  One finds \cite{Berrehrah:2014kba}

{\setlength\arraycolsep{0pt}
\begin{eqnarray}
\label{equ:Sec5.2}
& & \frac{d E^{\textrm{off}}}{d t} (p, T, \mu_q) = \frac{4}{(2 \pi)^7} \ \Pi_{i \in{p,q,p',q'}} \int m_p  m_i d m_i \ \rho_i^{WB} (m_i) \ \int_0^{\infty} \frac{q^4 m_1^{\textrm{off}} (s)}{s^2 E_q} \ \Biggl[- q f_0 (q) + \frac{p}{E} (M_Q + E_q) f_1 (q) \Biggr] \ d q,
%\nonumber
\end{eqnarray}}
 where $m_1^{\textrm{on}} (s)$, $m_1^{\textrm{off}} (s)$ and $f_1 (q)$ in (\ref{equ:Sec5.1}) and (\ref{equ:Sec5.2}) are given by
{\setlength\arraycolsep{0pt}
\begin{eqnarray}
\label{equ:Sec5.3}
& & m_1^{\textrm{on}} (s) := \frac{1}{8 p_{cm}^4} \int_{-4 p_{cm}^2}^0 \ \frac{1}{g_p g_Q} \sum_{i,j} \sum_{k,l} |\mathcal{M}|^2 (s, t; i, j | k, l) \times (- t) d t.
\nonumber\\
& & {} m_1^{\textrm{off}} (s) := \frac{1}{4 p^i p^f} \int_{t_{min}}^{t_{max}} \ \frac{1}{g_p g_Q} \sum_{i,j} \sum_{k,l} |\mathcal{M}_{2,2}^{\textrm{off}} (s, t; i, j | k, l)|^2 \ \Biggl[1 - \frac{t - (M_Q^i)^2 + (M_Q^f)^2 - 2 E^i E^f}{2 p^i p^f} \Biggr] \ d t,
\nonumber\\
& & {} f_n (q) = \frac{1}{2} \int d \cos \theta_{r} \ f \left(\frac{u^0 E_q - uq \cos \theta_{r} \mp \mu_q}{T} \right) \cos^n \theta_{r} \ .
\end{eqnarray}}

\begin{figure}[h!] %tbh!
\begin{center}
   \begin{minipage}[c]{.55\linewidth}
     \includegraphics[height=8.2cm, width=8.8cm]{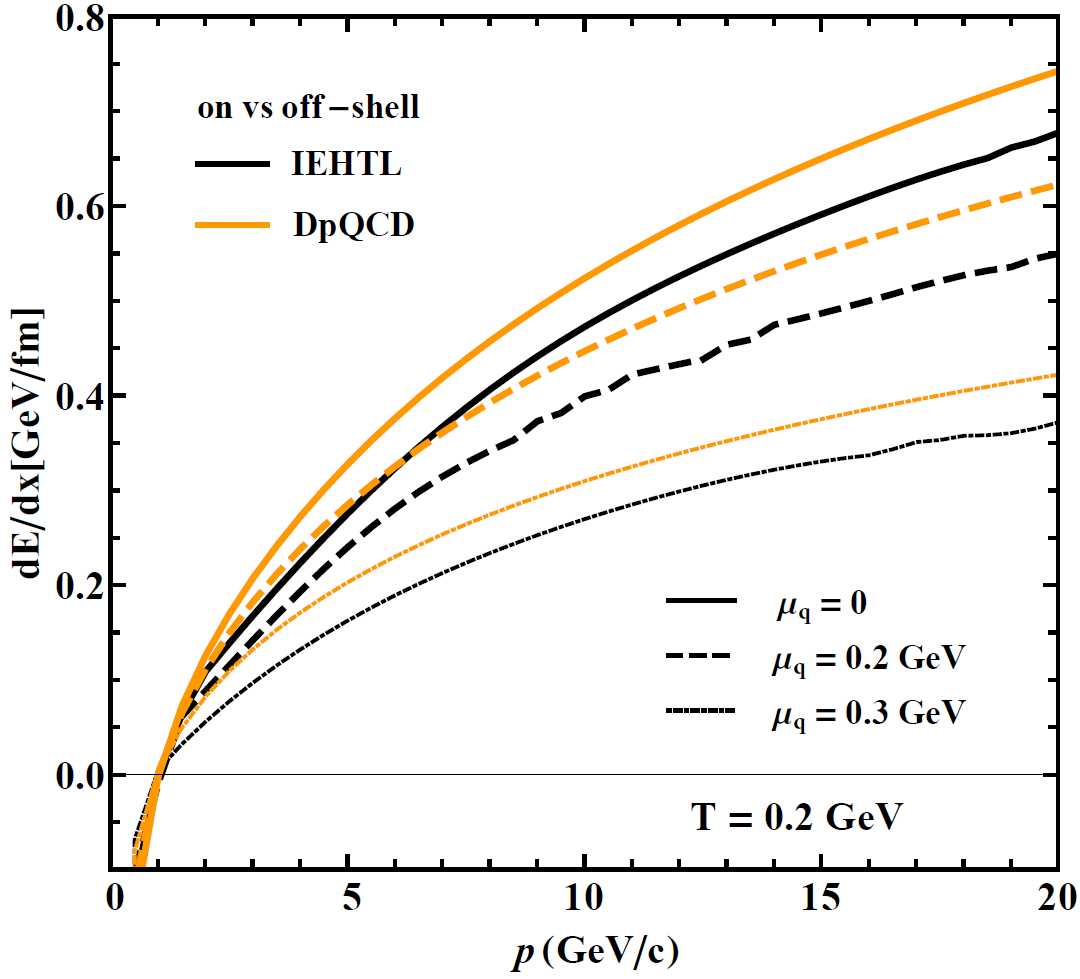}
   \end{minipage} \hfill
   \begin{minipage}[c]{.4\linewidth}
\caption{\emph{(Color online) The $c$-quark energy loss, $dE/dx$, in the plasma rest frame from the DpQCD and IEHTL models as a function of the heavy-quark momentum $p$ at $T = 0.2$ GeV  for different values of the quark chemical potential $\mu_q$.}\label{fig:cdEdxpQ}}
\end{minipage}
\end{center}
\end{figure}

  Furthermore, $p^i$ ($M_Q^i$) is the initial heavy momentum (mass) and $p^f$ ($M_Q^f$) is the final heavy momentum (mass). The heavy-quark energy loss (eq.\ref{equ:Sec5.1}) and (eq.\ref{equ:Sec5.2}) is illustrated in Fig. \ref{fig:cdEdxpQ} as a function of the heavy quark momentum for the DpQCD and IEHTL models at $T=0.2$ GeV for different  quark chemical potentials up to $\mu_q$ = 0.3 GeV. As for the rates, the off-shell spectral function decreases the energy loss as compared to the on-shell case as described by the DpQCD model. The difference is uniform as a function of the heavy quark momentum $p$ and the increase of the quark chemical potential. For low heavy-quark momenta and higher medium temperatures, the heavy quark looses less energy by elastic collisions. At low momentum heavy quarks start to gain energy to approach thermal equilibrium and to arrive at the average energy in the heat bath ($dE\!/\!dx$ is negative). For a medium at the temperature $T=0.2$ GeV, the increase of $\mu_q$ leads to a decrease of the heavy-quark energy loss. Nevertheless, varying both the medium temperature and $\mu_q$ influences the heavy-quark energy loss in different ways.

\begin{figure}[h!] %tbh!
\begin{center}
   \begin{minipage}[c]{.495\linewidth}
     \includegraphics[height=8cm, width=9cm]{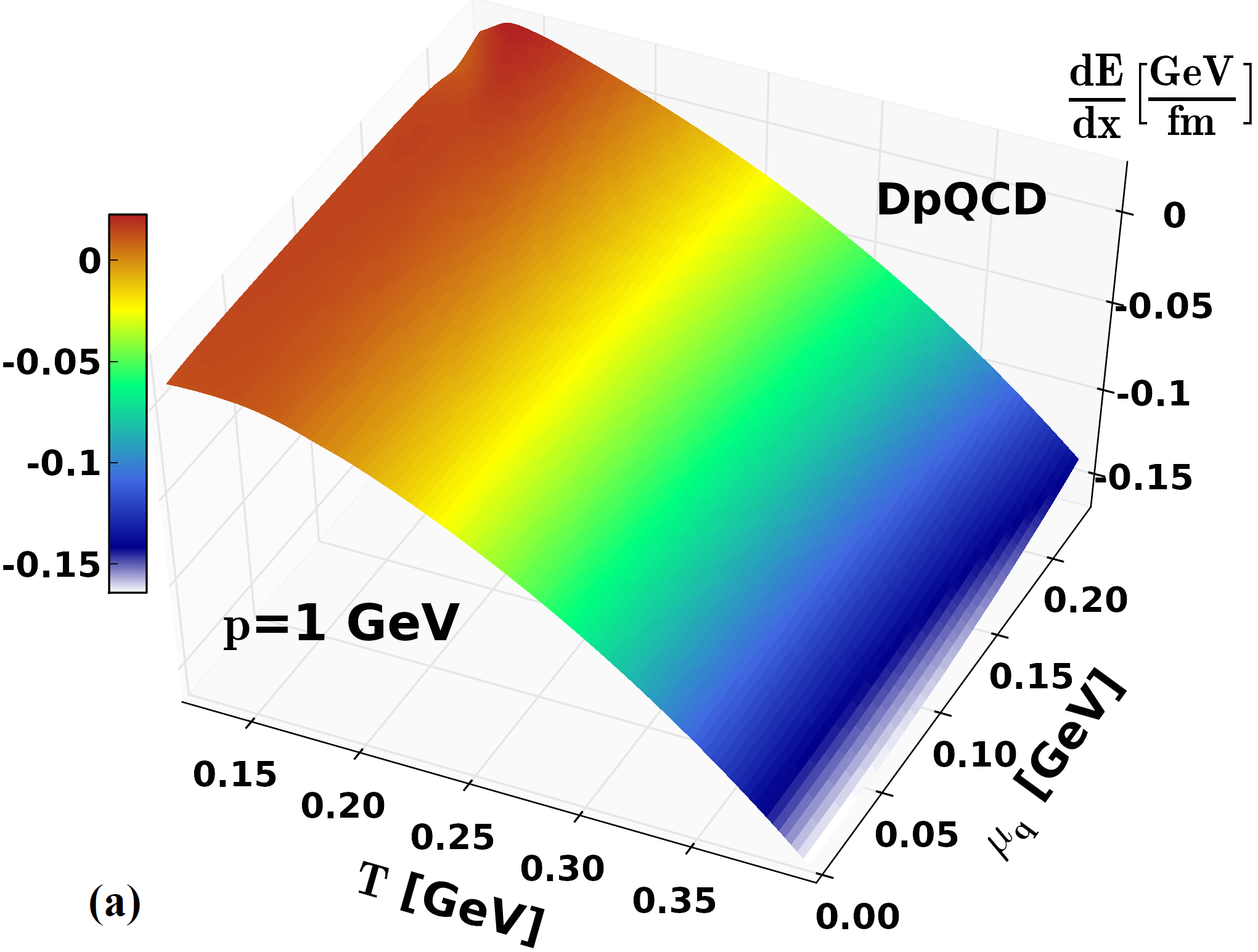}
   \end{minipage} \hfill
   \begin{minipage}[c]{.495\linewidth}
     \includegraphics[height=8cm, width=9cm]{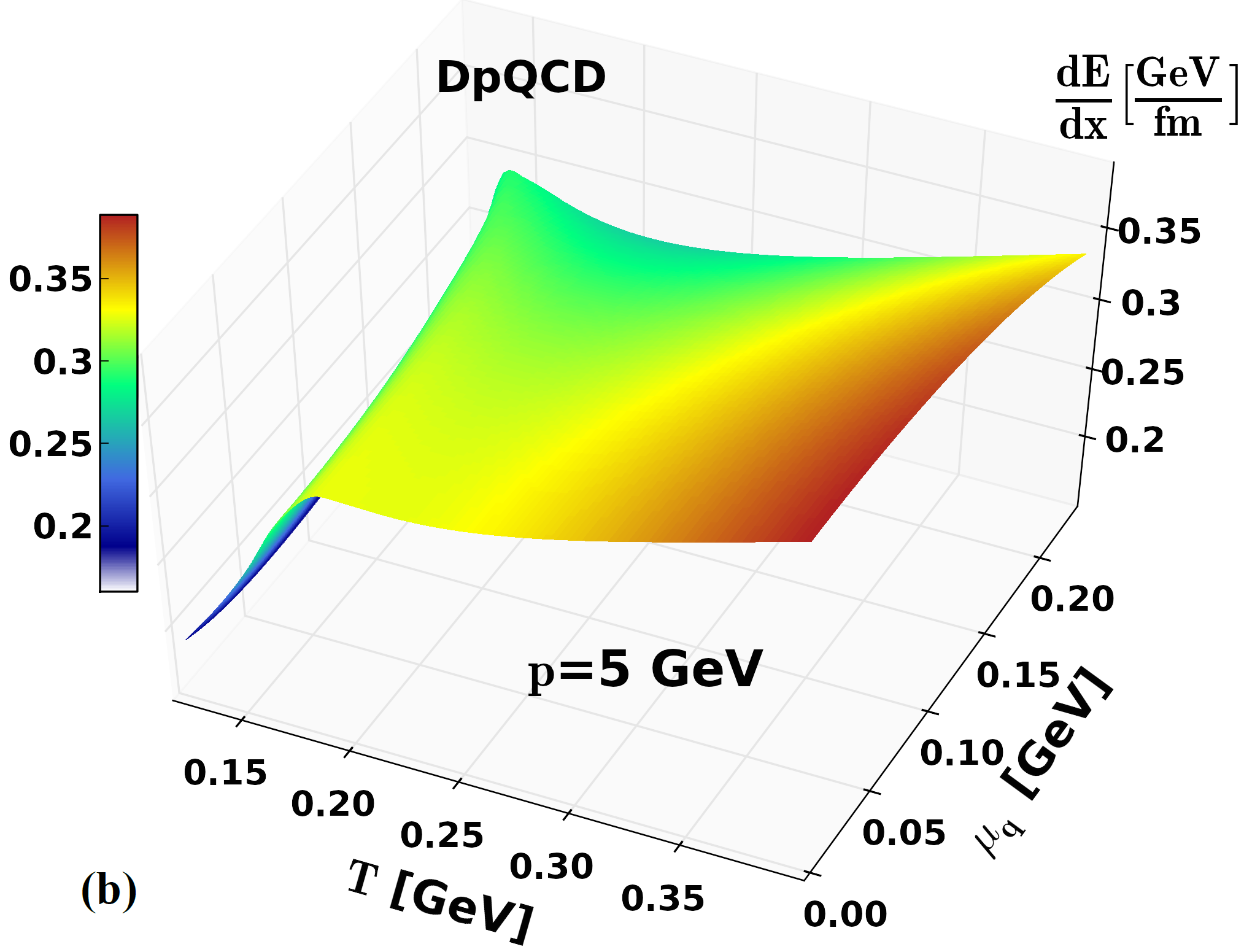}
\end{minipage}
\caption{\emph{(Color online) The $c$-quark energy loss, $dE/dx$, in the plasma rest frame from the DpQCD approach as a function of the temperature $T$ and quark chemical potential $\mu_q$ for the heavy-quark momentum $p = 1$ GeV (a) and $p = 5$ GeV (b).}}
\label{fig:cdEdxTmu}
\end{center}
\end{figure}

 The temperature and quark chemical potential dependencies of the collisional energy loss for low ($p = 1$ GeV/c) and intermediate heavy-quark  momentum ($p = 5$ GeV/c) is displayed in Fig. \ref{fig:cdEdxTmu}-(a) and (b) for the on-shell DpQCD model, respectively. Figs. \ref{fig:cdEdxTmu}- (a) and (b) show that the energy loss is increasing as a function of $p$. For intermediate momenta $p$ the conclusions drawn for the rate are valid also for the collisional energy loss; especially Fig. \ref{fig:cdEdxTmu}-(b) shows that with increasing temperature for small values of $\mu_q$ -- which corresponds to the increase of the parton masses in the DpQCD-- the energy loss becomes larger. This is more pronounced for large heavy-quark momenta. On the contrary, for $p = 1$ GeV (cf. Fig \ref{fig:cdEdxTmu}-a), where we observe an energy gain (a negative energy loss), the energy gain increases with the distance of the energy of the heavy quark from its equilibrium value. Nevertheless, as expected for a cross-over transition (at finite but small values of $\mu_q$) we observe a very smooth dependence of the energy loss on both variables, $T$ and $\mu_q$. The dependencies of $dE/dx$ on $T$ and $\mu_q$ are studied in \cite{Vija:1994is} using an extension of finite temperature HTL calculation to include finite $\mu_q$ in the distribution functions of the quark/anti-quarks. The effect of finite $\mu_q$ on $dE/dx$ seen in \cite{Vija:1994is} is much smaller than in our DpQCD and IEHTL models, shown in Fig. \ref{fig:cdEdxpQ}. We note that the finite $\mu_q$ effects in both the parton masses, coupling constant and distribution function are included in our model, which give to our calculations more inclusive effects than those of \cite{Vija:1994is}.

\begin{figure}[h!] %tbh!
\begin{center}
     \includegraphics[height=6.43cm, width=5.877cm]{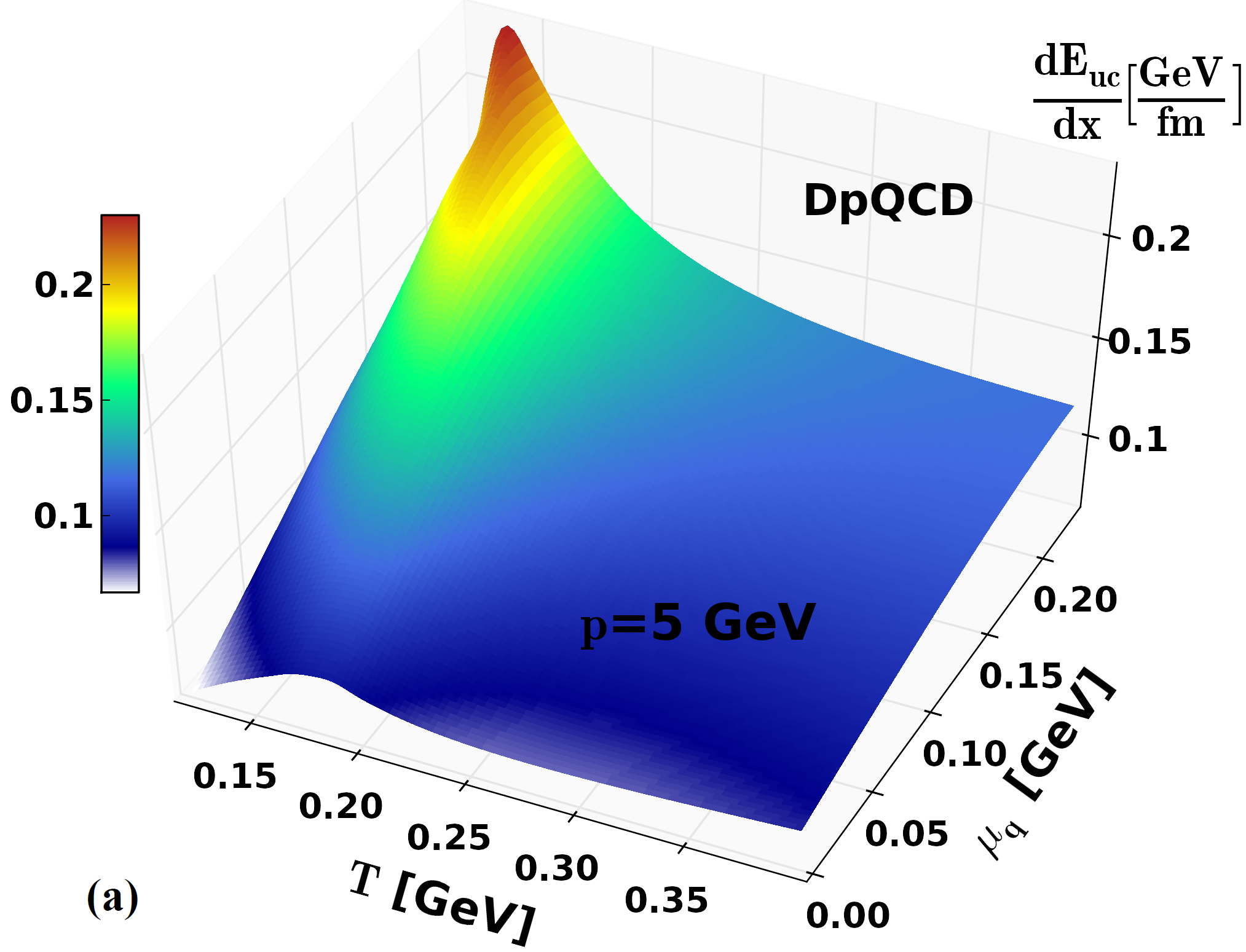}
     \hspace{0.05cm}
     \includegraphics[height=6.43cm, width=5.877cm]{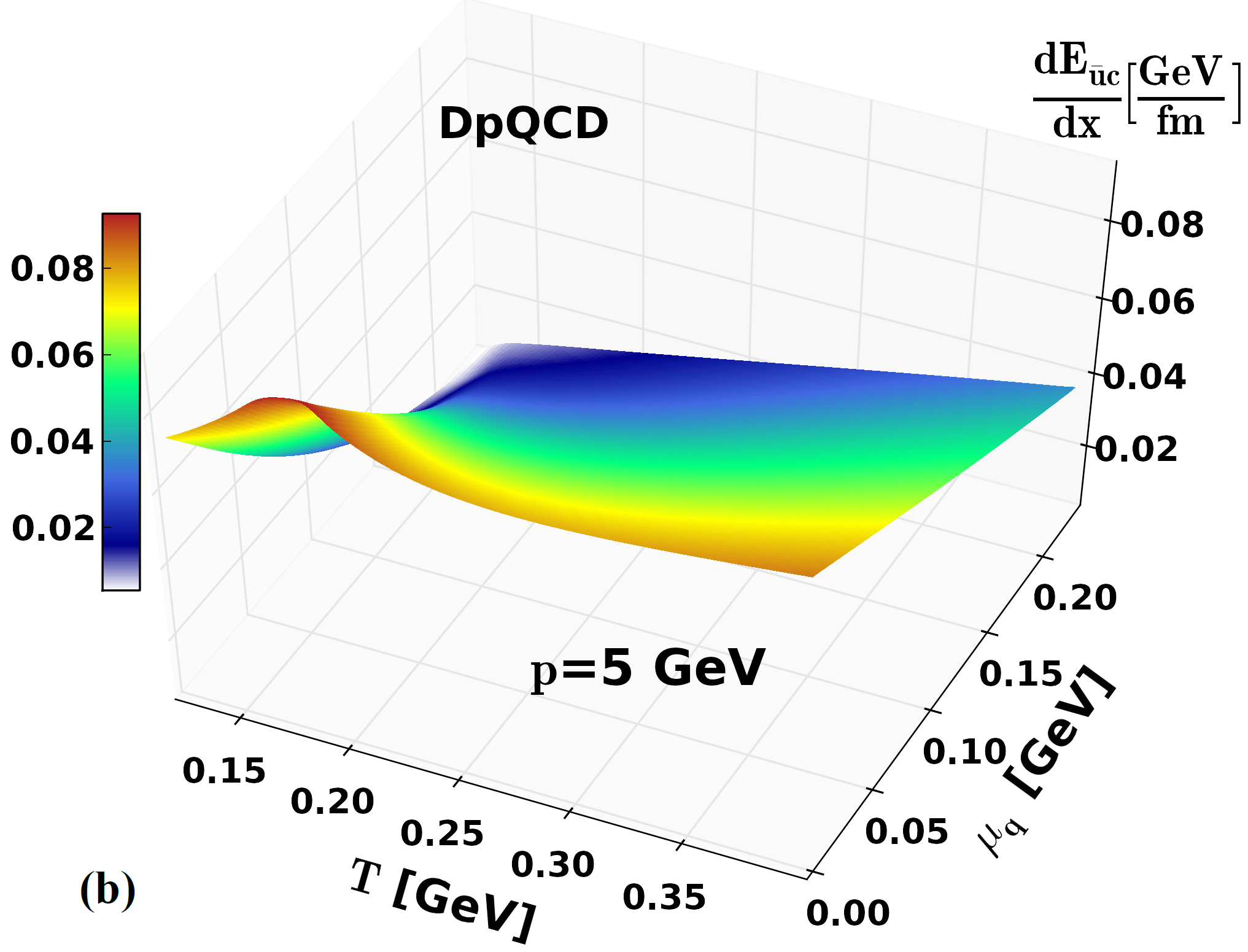}
     \hspace{0.05cm}
     \includegraphics[height=6.43cm, width=5.877cm]{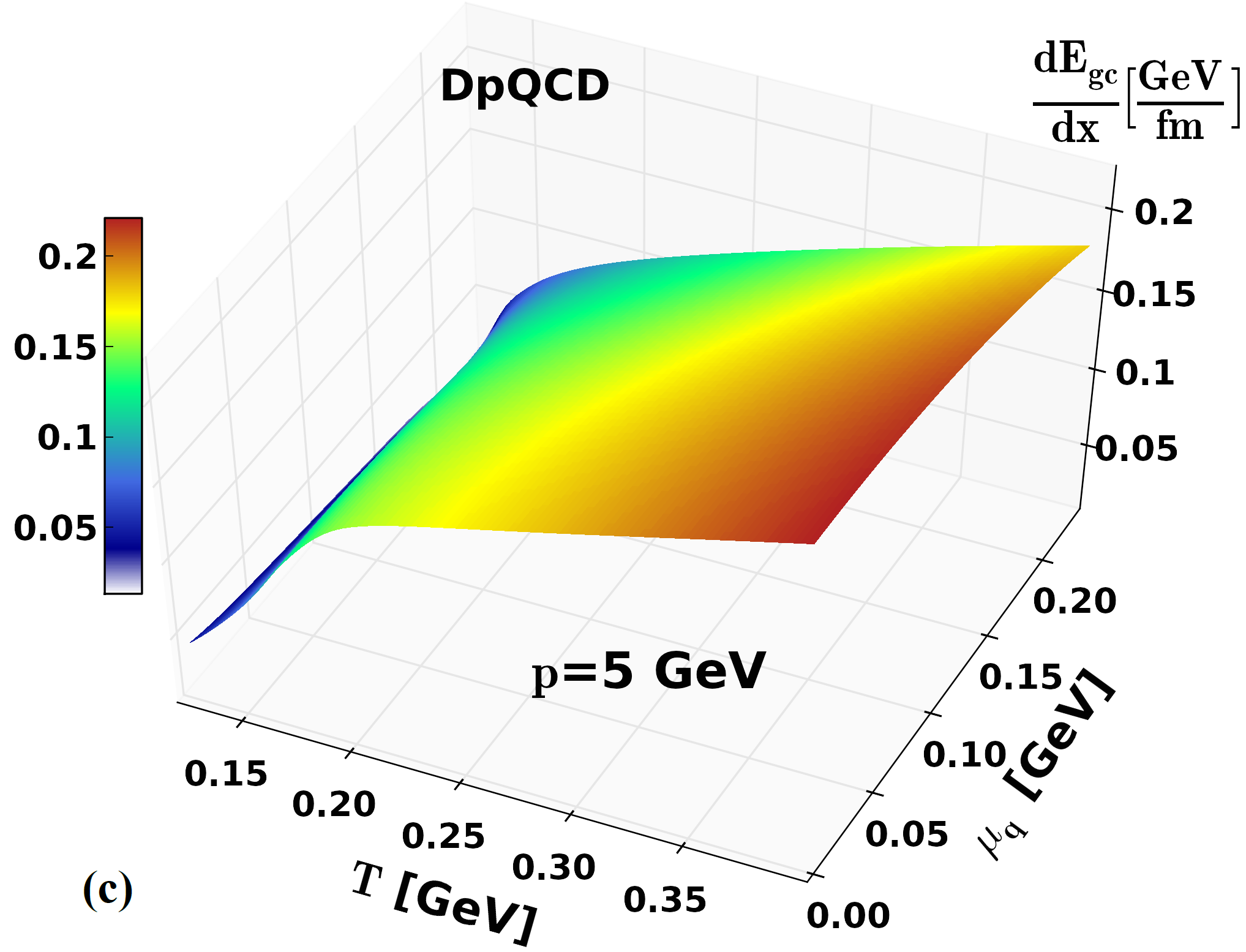}
%\vspace{-0.1cm}
\caption{\emph{(Color online)  The energy loss $dE/dx$ of c-quarks in the plasma rest frame as a function of the temperature $T$ and the quark chemical potential $\mu_q$ due to the scattering with light quarks (a), light antiquarks (b) and gluons (c). The on-shell heavy quark momentum in this case is $p = 5$ GeV.}}
\label{fig:cdEdxvsTmu}
\end{center}
\end{figure}

   The energy loss of a heavy quark due to its scattering with the QGP partons varies with the abundance of the particle species. Whereas the scattering of heavy quarks on the gluons is slightly $\mu_q$ dependent, the scattering on light quarks/antiquarks induces an energy loss which is highly ($T$, $\mu_q$) dependent. Figs. \ref{fig:cdEdxvsTmu}-(a), (b) and (c) illustrate this dependence of the heavy-quark collisional energy loss with quarks, anti-quarks and gluons of a finite temperature $T$ and quark chemical potential $\mu_q$ medium for an intermediate heavy-quark momentum ($p = 5$ GeV). As seen for the interaction rate, the profiles observed for the $q(\bar{q})-Q$ collisional energy loss evolve  oppositely due to the Fermi-Dirac distribution. As pointed out before, the quark (antiquark) density changes substantially with $\mu_q$.

  %---------------------------------------------------------------------------------------------------------------------------------------------
\section{Heavy quark momentum loss in a medium at finite temperature and chemical potential}
\label{HQPLoss}
%---------------------------------------------------------------------------------------------------------------------------------------------
\vskip -2mm

  We continue our study with the momentum loss of heavy quarks in a partonic medium at finite temperature and chemical potential. The drag ($A$) and diffusion ($B$) coefficients are evaluated according to \cite{Svetitsky:1987gq,GolamMustafa:1997id,Gossiaux:2004qw} by a Kramers-Moyal power expansion of the collision integral kernel of the Boltzmann equation. Note that the diffusion tensor $B$ admits a transverse-longitudinal decomposition (perpendicular and along the direction of the heavy quark in the fluid rest frame) and
contains two independent coefficient $B_{\parallel}$ and $B_{\perp}$.

\subsection{Longitudinal momentum loss} %-------------------------------------------------------------------------------------------------------------------------------------

 The drag coefficient $A$ describes the time evolution of the average of the longitudinal component of the momentum transfer $(p - p')^l$ of the heavy quark. For on-shell partons it has been defined in the plasma rest frame by Svetitsky \cite{Svetitsky:1987gq,GolamMustafa:1997id} and evaluated in \cite{Berrehrah:2014kba} by

{\setlength\arraycolsep{-6pt}
\begin{eqnarray}
\label{equ:Sec6.1}
A^{\textrm{on}} (p, T, \mu_q) = \frac{d<\!\!p\!\!>}{dt} = \frac{M_Q}{4 (2 \pi)^3} \int_0^{\infty} \frac{q^4 m_1^{\textrm{on}} (s)}{s^2 E_q} \ \Biggl[(M_Q + E_q) f_1 (q) - \frac{p}{E} q f_0 (q) \Biggr] \ d q.
%\nonumber\\
%& & {}
\end{eqnarray}}
  Eq.(\ref{equ:Sec6.1}) is easily extended for the off-shell case. One finds \cite{Berrehrah:2014kba},

{\setlength\arraycolsep{0pt}
\begin{eqnarray}
\label{equ:Sec6.2}
& & \hspace{-1cm} A^{\textrm{off}} (p, T, \mu_q) = \frac{d<\!\!p\!\!>}{dt} = \frac{4}{(2 \pi)^7} \ \Pi_{i \in{p,q,p',q'}} \int m_p  m_i d m_i \ \rho_i^{BW} (m_i) \ \int_0^{\infty} \frac{q^4 m_1^{\textrm{off}} (s)}{s^2 E_q} \ \Biggl[(M_Q + E_q) f_1 (q) - \frac{p}{E} q f_0 (q) \Biggr] \ d q,
%\nonumber
\end{eqnarray}}
 where $m_1^{\textrm{on}} (s)$, $m_1^{\textrm{off}} (s)$ and $f_{0,1} (q)$ are given in (\ref{equ:Sec5.3}).

\subsection{Transverse momentum loss} %-------------------------------------------------------------------------------------------------------------------------------------

 During its propagation through the plasma a heavy quark receives random kicks (momentum transfers) from the constituents of the medium. The average value of the transverse momentum remains zero but the variance $<p_{\perp}^2>$ increases with the number of collisions. The increase of the variance $<\!p_{T}^2\!>$ per unit length is called the transport coefficient $\hat{q}$ and defined as 
 \begin{equation} \label{qhat}  \hat{q} = \frac{d <\!p_{\perp}^2\!>}{dx} = \frac{<\!p_{\perp}^2\!>_{single\ coll}}{\ell} .\end{equation} Using the relation between the transverse second moment $B_{T}$ and $\hat{q}$ given by $\hat{q} = \frac{4 E}{p} \ B_{T}$, we can evaluate $\hat{q}$ for massive on-shell partons (as in \cite{Berrehrah:2014kba}) by the expression:

{\setlength\arraycolsep{-6pt}
\begin{eqnarray}
\label{equ:Sec6.3}
& & \hat{q}^{\textrm{on}} (p, T, \mu_q) = \frac{M_Q^3}{2(2\pi)^3 p} \int_{0}^{+ \infty} \frac{q^5}{s^2 E_q} \Biggl[\frac{m_1^{\textrm{on}} (s) - m_2^{\textrm{on}} (s)}{2} \left(f_{0} + f_{2} \right) + \frac{(E_q + M_Q)^2}{s} m_2^{\textrm{on}} (s) \left(f_{0} - f_{2} \right) \Biggr] \ d q,
%\nonumber\\
%& & {}
\end{eqnarray}}
 where $m_1^{\textrm{on}} (s)$ and $f_{0,1,2} (q)$ are given in (\ref{equ:Sec5.3}) and $m_2^{\textrm{on}} (s)$ by

{\setlength\arraycolsep{-10pt}
\begin{eqnarray}
\label{equ:Sec6.4}
\hspace*{-0.5cm} m_2^{\textrm{on}} (s) = \int_{-1}^1 d \cos \theta  \left(\frac{1 - \cos \theta}{2} \right)^2 \frac{1}{g_g g_Q} \sum_{i j} \sum_{k l} |\mathcal{M}^{\textrm{on}}|^2 (s, t; i j|l, k) = \frac{1}{32 p_{cm}^6} \int_{- 4 p_{cm}^2}^{0} \ \frac{1}{g_g g_Q} \sum_{i j} \sum_{k l} |\mathcal{M}^{\textrm{on}}|^2 (s, t; i j|l, k) \ t^2 d t.
%\nonumber\\
%& & {}
\end{eqnarray}}
  The off-shell transport coefficient $\hat{q}^{\textrm{off}}$ is deduced by using  Breit-Wigner spectral functions for the off-shell partons and extending the on-shell $\hat{q}^{\textrm{on}}$ as

{\setlength\arraycolsep{-10pt}
\begin{eqnarray}
\label{equ:Sec6.5}
& & \hspace*{-1.2cm} \hat{q}^{\textrm{off}} (p, T, \mu_q) \!=\! \frac{1}{2 (2 \pi)^3} \Pi_{i \in{p,q,p',q'}} \!\!\!\! \int \frac{m_p^2}{p} m_i d m_i \rho_p^{BW} (m_i) \!\!\!\! \int \frac{q^5 m_p}{s^2 E_q} \!\! \Biggl[\!\frac{m_1^{\textrm{off}} (s) - m_2^{\textrm{off}} (s)}{2} \left(f_{0} + f_{2} \right) + \frac{(E_q + m_p)^2}{s} m_2^{\textrm{off}} (s) \left(f_{0} - f_{2} \right) \!\!\Biggr] \! d q
%\nonumber\\
%& & {}
\end{eqnarray}}
with:
{\setlength\arraycolsep{-6pt}
\begin{eqnarray}
\label{equ:Sec6.6}
m_2^{\textrm{off}} (s) = \frac{1}{8 p^i p^f} \int_{t_{min}}^{t_{max}} \ \frac{1}{g_g g_Q} \sum_{i j} \sum_{k l} |\mathcal{M}^{\textrm{off}}|^2 (s, t; i, j|l, k) \Biggl[1 - \frac{t - (M_Q^i)^2 + (M_Q^f)^2 - 2 E^i E^f}{2 p^i p^f} \Biggr]^2 d t.
%\nonumber
\end{eqnarray}}

\subsection{Numerical results} %--------------------------------------------------------------------------------------------------------------------------------------------

\begin{figure}[h!] %tbh!
\begin{center}
   \begin{minipage}[c]{.49\linewidth}
     \includegraphics[height=8.7cm, width=8.9cm]{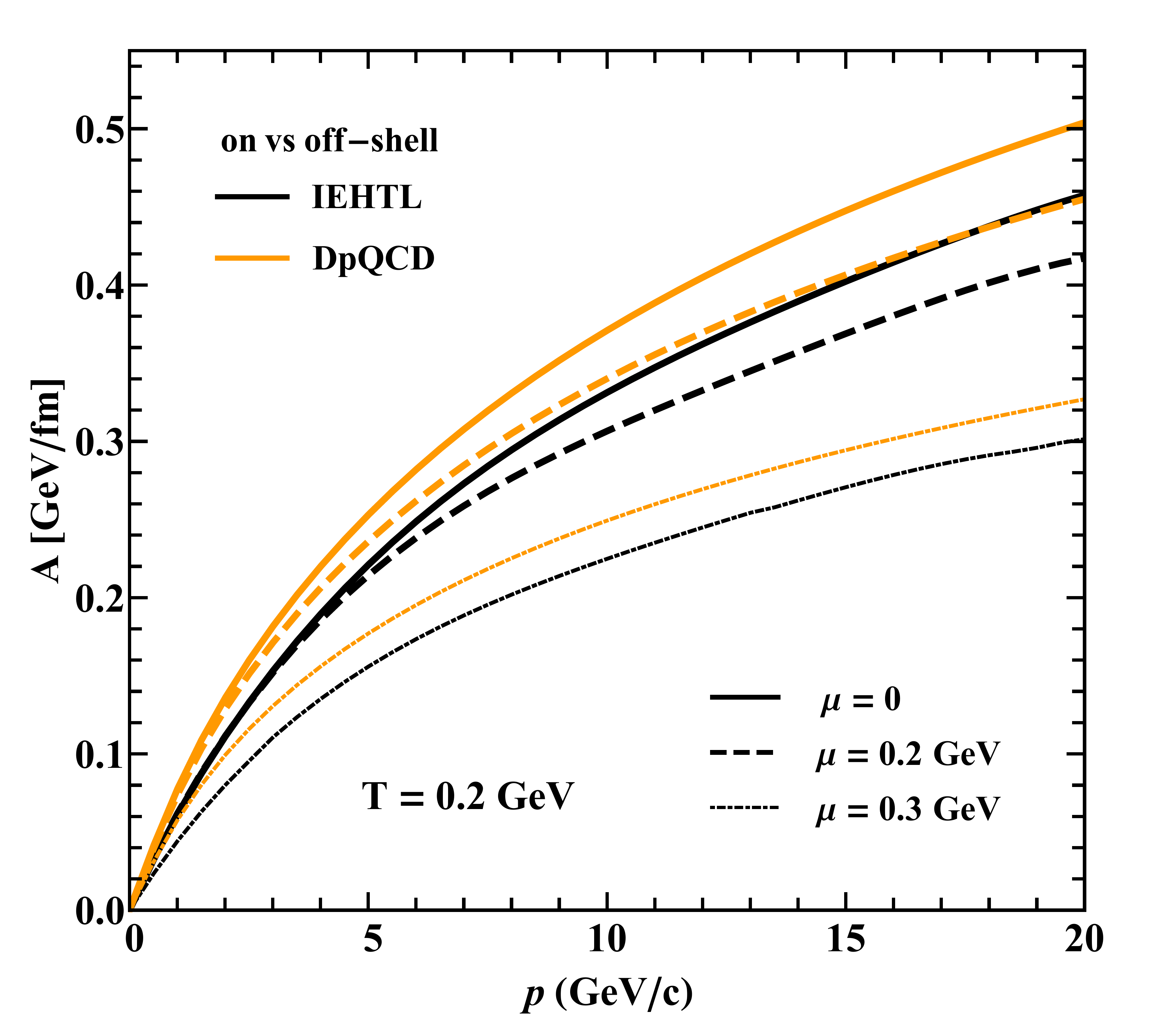}
   \end{minipage} \hfill
   \begin{minipage}[c]{.49\linewidth}
     \includegraphics[height=8.7cm, width=8.9cm]{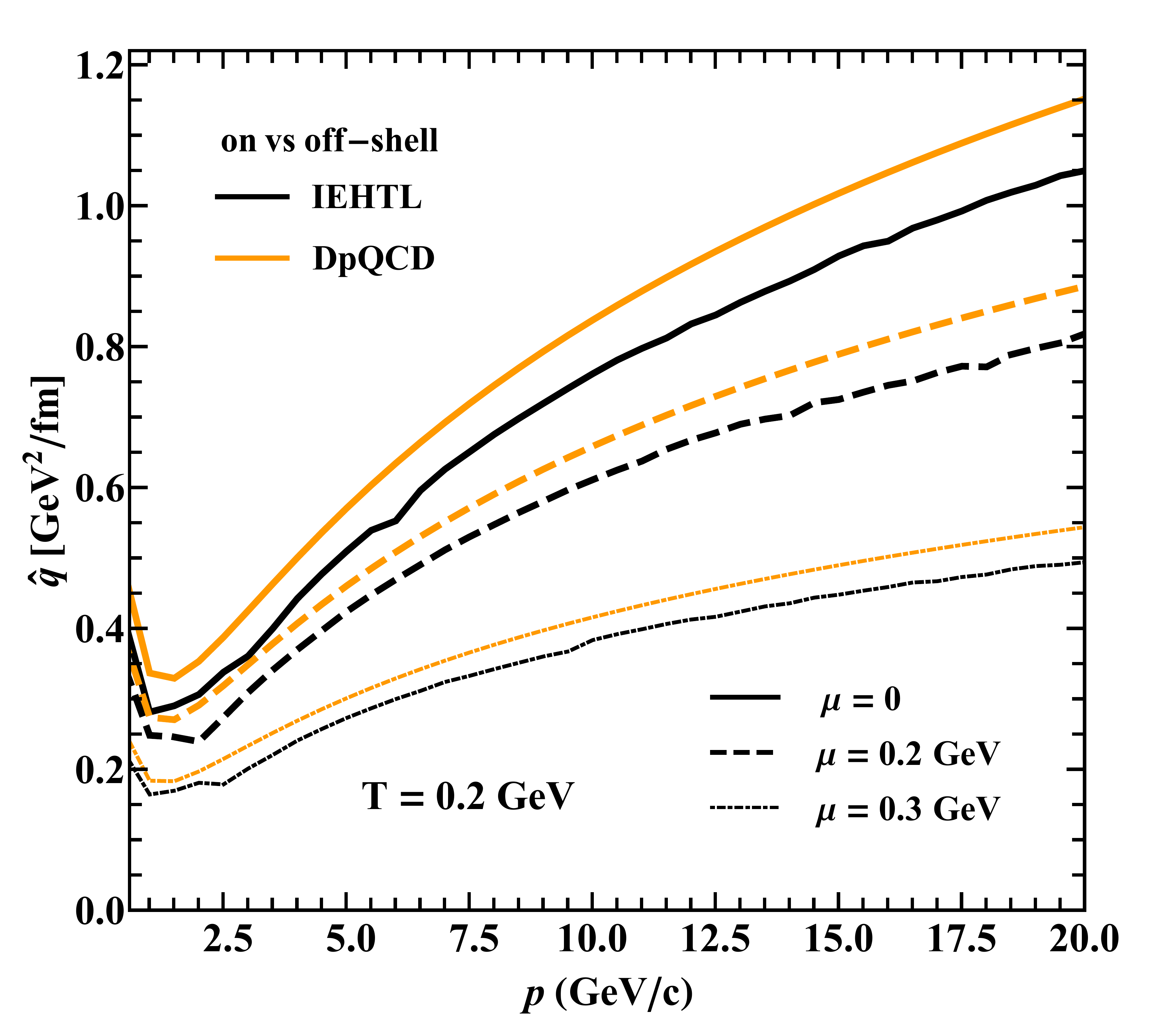}
\end{minipage}
\vspace{-0.2cm}
\caption{\emph{(Color online) $c$ quark drag coefficient, (\ref{equ:Sec6.1}) and (\ref{equ:Sec6.2}), (a) and $\hat{q}$, (\ref{equ:Sec6.3}) and (\ref{equ:Sec6.5}), (b) in the plasma rest frame as a function of the heavy quark momentum $p$ for $T = 2$ GeV.}}
\label{fig:cDragqhat}
\end{center}
\end{figure}

  We discuss now the drag coefficient and the transport coefficient $\hat{q}$ for the models DpQCD and IEHTL from Ref. \cite{Berrehrah:2013mua} at finite $T$ and $\mu_q$. In the DpQCD approach, the light and heavy quark and gluon masses are given by the DQPM pole masses. Fig. \ref{fig:cDragqhat}-(a) shows the drag coefficient (\ref{equ:Sec6.1}) and (\ref{equ:Sec6.2}) of heavy quarks as a function of the heavy quark momentum $p$. The temperature of the heath bath is chosen as $T=0.2$ GeV. Figure \ref{fig:cDragqhat}-(b) illustrates the influence of a finite parton width on the heavy quark transport coefficient $\hat{q}$, where $\hat{q}$ is displayed for on-shell partons (DpQCD) and off-shell partons (IEHTL) as a function of the heavy quark momentum and for a temperature of $T=0.2$ GeV and different values of $\mu_q$. For both $A$ and $\hat{q}$ the off-shell partons case is slightly lower by about 20\% independent of momentum, temperature or quark chemical potential than for the corresponding on-shell case.

\begin{figure}[h!] %tbh!
\begin{center}
   \begin{minipage}[c]{.495\linewidth}
     \includegraphics[height=8.3cm, width=8.9cm]{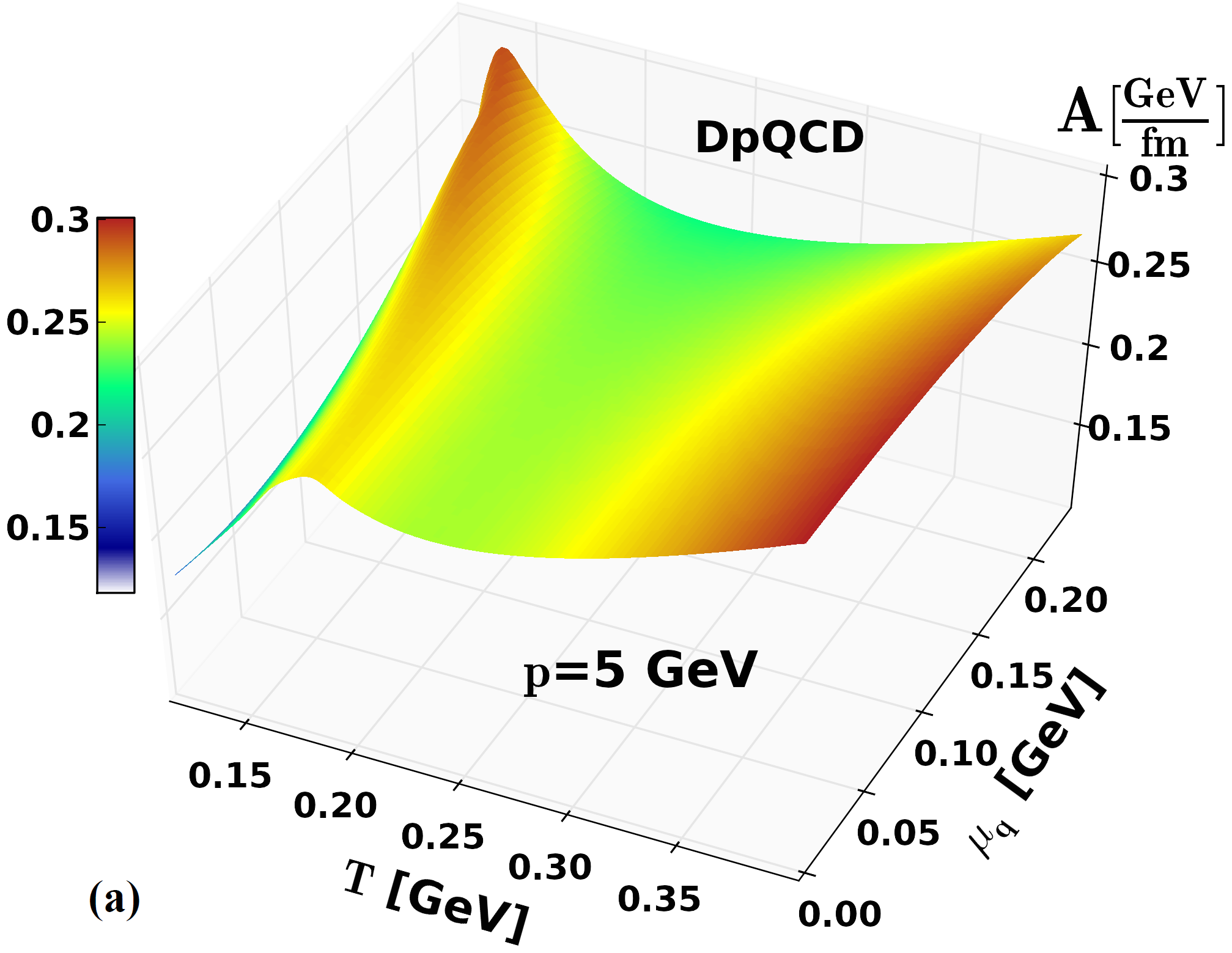}
   \end{minipage} \hfill
   \begin{minipage}[c]{.495\linewidth}
     \includegraphics[height=8.3cm, width=8.9cm]{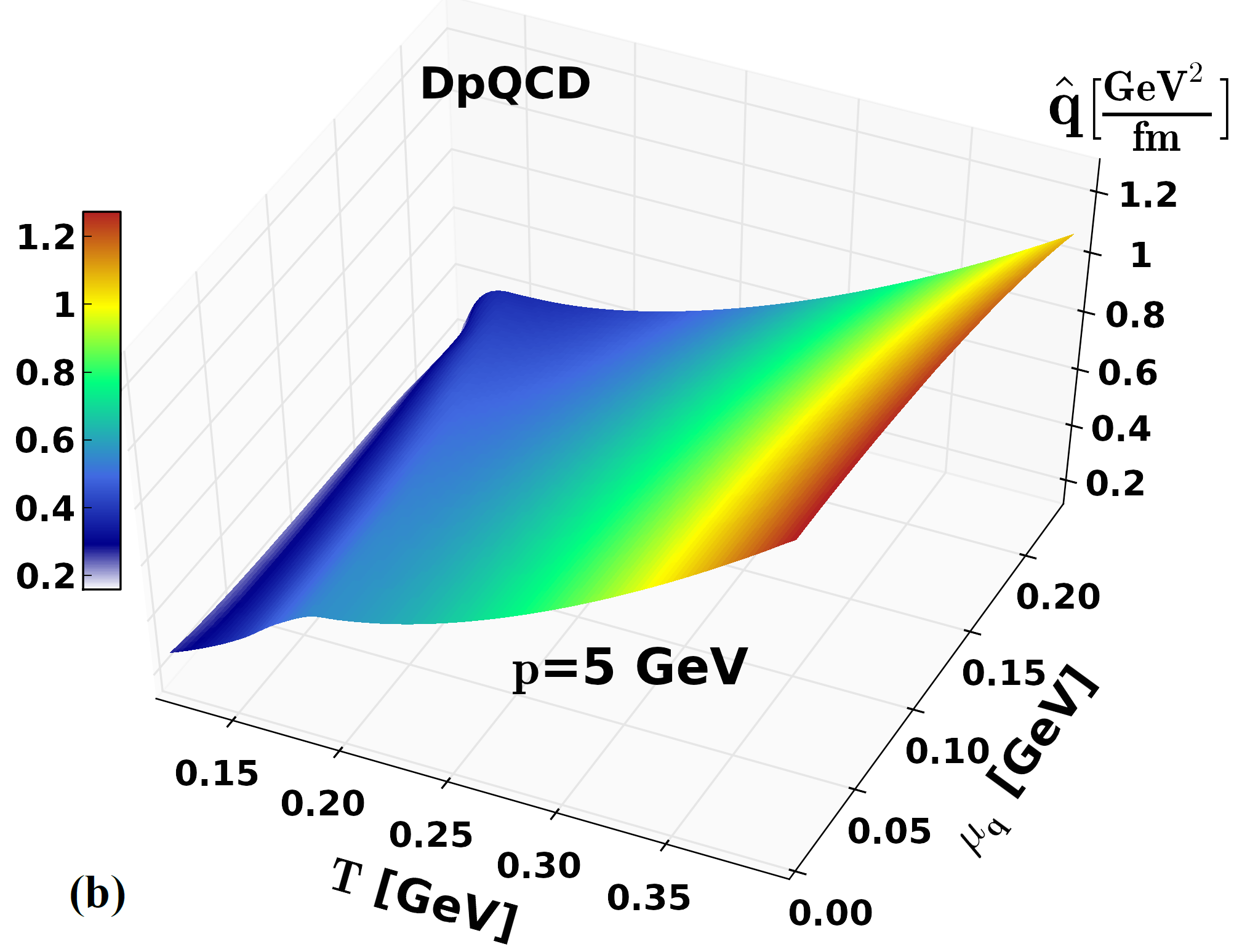}
\end{minipage}
\caption{\emph{(Color online) $c$-quark drag (a) and $\hat{q}$ (b) in the plasma rest frame from the DpQCD approach as a function of the temperature $T$ and quark chemical potential $\mu_q$ for the heavy-quark momentum $p = 5$ GeV.}}
\label{fig:cAqhatTmu}
\end{center}
\end{figure}

   The temperature and quark chemical potential dependences of the drag coefficient and $\hat{q}$ for intermediate heavy quark  momentum ($p = 5$ GeV/c) is displayed in Fig. \ref{fig:cAqhatTmu}-(a) and (b) for the on-shell DpQCD model, respectively. Fig. \ref{fig:cAqhatTmu}- (a) and (b) shows that both the longitudinal and transverse momentum is increasing for high temperatures and low $\mu_q$.  On the contrary, for low temperatures and large $\mu_q$ the longitudinal momentum loss is still considerable whereas the transverse momentum transfer is low. We conclude that longitudinal momentum transfers are important not only in a hot medium but also in a dense medium whereas the dense medium leads to less transverse fluctuations in the heavy quark propagation. The relatively large drag at low temperatures in DpQCD/IEHTL is due to the strong increase of the running coupling $\alpha_s (T, \mu_q)$ (infrared enhancement) for temperatures close to $T_c (\mu_q)$.

   Regarding the variations in $T$ and $\mu_q$ of the longitudinal and transverse momentum losses, one may roughly presage the tendency of the nuclear modification factor $R_{AA}$ and the elliptic flow of charm particles in a hot and dense medium. We expect that the $R_{AA}$ of charmed mesons will still be small in a hot and dense medium as in case of finite temperature and zero chemical potential. On the contrary, the elliptic flow $v_2$ is expected to be much smaller in a hot and dense medium compared to our knowledge from RHIC and LHC measurements for $\mu_q \approx 0$, since the dense medium damps the transverse momentum fluctuations. Moreover, one could notice that these finite $\mu_q$ calculations might have interesting consequences on the $c$-quark angular correlations at FAIR, which should be more peaked than those seen at RHIC or LHC. However, such predictions have to be confirmed by explicit transport simulations of the finite temperature and chemical potential QGP, since the macroscopic observables ($R_{AA}$ and $v_{2}$) depend not only on the microscopic processes and transport coefficients, as studied in this work, but also on the time evolution of the QGP. Such studies will be carried out in the near future using the Parton-Hardon-String Dynamics (PHSD) transport approach \cite{Cassing:2009vt,Cassing:2011vt} for heavy-ion collisions from $\sqrt{s_{NN}} =$ 5 GeV to 10 GeV \cite{Vadim}. 
  \section{Summary}
\label{summary}
\vskip -2mm

  We have presented in this work on- and off-shell approaches to describe the microscopic interactions between a heavy quark and the QGP degrees of freedom in a partonic medium at finite temperature $T$ and quark chemical potential $\mu_q$. In our formalism  both the perturbative and non-perturbative parts of QCD are involved (DpQCD approach) and an off-shell description of heavy-quark interactions in the QGP (IEHTL) at finite T and $\mu_q$. For each of these models, we have calculated the differential cross sections and confronted their implications on the usual mesoscopic observables (energy loss, drag and diffusion coefficients, longitudinal and transverse $Q$ momentum fluctuations, etc.) at finite $T$ and $\mu_q$.

  The fundamental parameters describing the heavy-quark collisional scattering have been fixed within the DQPM at finite temperature and chemical potential. The formulation of the DQPM at finite $\mu_q$ is based on the hypothesis that the phase boundary is close to the line of constant energy density in the $(T, \mu_q)$ plane which leads to the approximation (\ref{equ:Sec2.8}). On the other hand this approximation is in very good agreement with recent lQCD results \cite{Delia} on the expansion coefficient w.r.t. $\mu_q^2$. Accordingly we infer that the equation of state of QCD from $T_c(\mu_q)$ to higher temperatures is roughly under control at least for small/moderate quark chemical potentials.

  Our study demonstrates that even if the influence of the finite width of the quasi-particles on heavy quark scattering is small on the off-shell $Q$ scattering cross sections, a noticeable effect is seen in the off-shell transport coefficients (IEHTL model) as compared to the on-shell ones (DpQCD model). This is due to reduced kinematical thresholds in the off-shell cross sections and is independent of the variables ($T$, $\mu_q$).

  A medium in which the chemical potential is finite leads to a reduction of the $qQ$ and $gQ$ elastic cross section and consequently to a reduction of heavy-quark energy and momentum losses as compared to a zero $\mu_q$ medium. Nevertheless, we have concluded that longitudinal momentum transfers are important not only in a hot medium but also in a dense medium whereas the dense medium leads to less transverse fluctuations in the heavy quark propagation. The relative large drag at low temperatures in DpQCD/IEHTL is due to the strong increase of the running coupling $\alpha_s (T, \mu_q)$ (infrared enhancement) for temperatures close to $T_c (\mu_q)$.

  We have observed a smooth dependence of the energy loss on both variables $T$ and $\mu_q$ at finite but not too large values of $\mu_q$. Such a profile is expected for a cross-over transition from the partonic to the hadronic medium. For $\mu_q = 0$ the gluon mass depends on the temperature and therefore the increase of the energy loss is due to a change of the coupling. For $\mu_q = 0.2$ GeV, the energy loss is also increasing with temperature but less than for $\mu_q = 0$ because here both the coupling and the effective gluon mass decrease and the increase of the infrared regulator is counterbalanced by the decrease of the coupling. Since the variations of all transport coefficients with $T$ and $\mu_q$ are rather smooth (within the present DQPM propagators) the transition from hadronic degrees of freedom to partonic ones remains a crossover up to $\mu_q$ = 0.2 GeV. Present studies within the DQPM indicate a change to a first order transition only for $\mu_q >$ 0.3 GeV but this is yet model dependent and not robust.

  From the variations of the longitudinal and transverse momentum losses with $T$ and $\mu_q$ we expect a large suppression of $R_{AA}$ of charmed mesons in a hot and dense medium, but a much smaller value of the elliptic flow $v_2$ in a hot and dense medium as compared to our knowledge from RHIC and LHC for finite temperature and  approximately zero chemical potential. However, such expectations have to be confirmed by microscopic transport simulations (e.g. within PHSD) for heavy-ion collisions from AGS to SPS energies. This is also the energy regime of the future FAIR and NICA facilities and the BES program at RHIC that all address the properties of QCD at high baryon densities or high $\mu_q$, respectively.

  \vspace{-3mm}
\section*{Acknowledgment}
\vskip -3mm

 H. Berrehrah acknowledges the financial support through DFG and the ``HIC for FAIR'' framework of the ``LOEWE'' program. The computational resources have been provided by the LOEWE-CSC.

% The authors acknowledge valuable discussions with R. Marty and O. Linnyk, and the financial support through the `HIC for FAIR' framework of the `LOEWE' program.

% \input{appendix.tex}

  \bibliography{References}

\end{document}